\documentclass[prc,onecolumn,aps,floats,floatfix,showpacs,nofootinbib,%
superscriptaddress]{revtex4}

\usepackage{amsmath}
\usepackage{amssymb}
\usepackage{amsfonts}
\usepackage{graphicx}
\usepackage{tabularx}
\usepackage{multirow}
\usepackage{lscape}
\usepackage{rotating}
\usepackage{pstricks}
\usepackage{feynmf}
\usepackage{lscape}
\usepackage{feynmf}

\begin{document}

\title{Extrapolating from neural network models: a cautionary tale}

\author{A. Pastore}
\email{alessandro.pastore@york.ac.uk}
\affiliation{Department of Physics, University of York, Heslington, York, Y010 5DD, UK}

\author{M. Carnini}
\email{marco.carnini@features-analytics.com}
\affiliation{Features Analytics, Rue de Charleroi 2, 1400 Nivelles, Belgium}



 
\date{\today}
\begin{abstract}
We present three different methods to estimate error bars on the predictions made using a neural network. All of them represent lower bounds for the extrapolation errors. For example, we did not include an analysis on robustness against small perturbations of the input data.

At first, we illustrate the methods through a simple toy model, then, we apply them to some realistic cases related to nuclear masses. By using theoretical data simulated either with a liquid-drop model or a Skyrme energy density functional, we benchmark the extrapolation performance of the neural network in regions of the Segr\`e chart far away from the ones used for the training and validation.
Finally, we discuss how error bars can help identifying when the extrapolation becomes too uncertain and thus unreliable.\end{abstract}


\maketitle

\section{Introduction}

Neural networks~\cite{friedman2001elements} (NN) are powerful tools that are widely used in several domains of science. Within the nuclear physics community, several groups have started investigating the NN as a tool to improve current models in predicting specific observables such as nuclear masses~\cite{uta16,uta17,neufcourt2018bayesian,pastore2020impact} and radii~\cite{akkoyun2013artificial}, or as intermediate tool to avoid time-consuming calculations~\cite{lasseri2020taming}. 
The domain of application is so vast since NN are \emph{universal} approximators~\cite{cybenko1989approximation}: any continuous function can be approximated by a single-layer neural network with a sufficiently large number of neurons. Each  neuron forming the NN contains some adjustable (hyper-)parameters (weights and biases) that are determined in order to minimise a given \emph{objective} or \emph{loss} function, typically the mean squared error. 

Once the architecture of the network is fixed, \emph{i.e.} the number of layers,  nodes and the connection patterns, the critical aspect is to find the optimal set of data to optimise the values of the weights and biases. To some extent, this is the same procedure commonly used in nuclear physics to adjust model parameters~\cite{kortelainen2010nuclear,kortelainen2014nuclear}. As a consequence, it is important to equip the NN with a reasonable estimate of the error bars to help assessing the quality of the results.

As discussed in Ref.\cite{dobaczewski2014error}, estimating theoretical errors is not an easy task since many factors could contribute to them. In particular, one can model the prediction error in terms of two main components: a \emph{statistical} and a \emph{systematic} one. The first arises from the optimisation procedure and it can be evaluated using specific statistical tools, while the second is usually unknown.

The standard strategy used to estimate error bars is based on the covariance matrix and the first derivative of the model in parameter space. See Refs.\cite{dobaczewski2014error,roca2015covariance} for more details.
Such an operation can not be applied to a typical NN for a very simple reason: the number of parameters is very large, typically in the range of thousands. Since the NN is non-linear in parameter space~\cite{barlow1989statistics}, it follows that the covariance matrix need to be evaluated by performing numerical derivatives in the parameter space. Due to the possible numerical issues discussed in Ref.~\cite{roca2015covariance}, we prefer not to explore this method.

In the present article, we study three different methods to estimate error bars that do not require major modification to the existing Python functions. To this purpose, we train a series of NNs using nuclear mass data derived from existing nuclear models: the liquid-drop~(LD)\cite{weizsacker1935theorie} and the Skyrme nuclear energy density function (NEDF)~\cite{bender2003self}. The idea is very simple: guided by the current knowledge of nuclear masses~\cite{wang2017ame2016}, we separate our theoretical masses in two sets: one corresponding to the measured ones ($\approx2400$ nuclei) and the other corresponding to the extrapolated region. 
This mimic the current situation when in the scientific literature several authors try to extrapolate models in regions where no data are available. The choice of using data generated through models and not experimental values is dictated by the desire of working on systems for which the values of the mass is available, to assess the validity of the error bars. We are aware that the trained NN will not produce meaningful estimations for nuclei. The best that can be achieved is to approximate the explicit formula of the model (say, Liquid Drop) through the NN.

Our goal it to show that even this simple approximation is challenging, and that extrapolation can be at best attempted close to the region of the known nuclear masses.

The article has been structured to be used as a guide to navigate the associated Jupiter notebooks made available as Supplementary material. The article is organised as follow: in Sec\ref{toymodel} we briefly introduce the NN and its error estimate and we apply it to a simple toy-model; in Sec.\ref{sec:nuclear} we present the nuclear models we use to obtain the data-set; in Sec.\ref{sec:error} we discuss the various methods to estimate error bars applied to realistic cases. Finally, we present our conclusions in Sec.\ref{sec:conc}.

\section{What is a Neural Network?}\label{toymodel}

A Neural Network~\cite{friedman2001elements} is an ensemble of elementary units called neurons arranged in layers and connected to each other. 
The individual neuron is represented mathematically as an \emph{activation} function $f$ (representing a neuron action potential), that takes some weighted inputs and sums them up to produce an output. In a formula, the output $y$ of a neuron is

\begin{eqnarray}
y=f(\mathbf{x}\cdot\mathbf{w}+b)\;,
\end{eqnarray}

\noindent  where $\mathbf{x}$ is the input vector, $\mathbf{w}$ is the vector of weights and $b$ the bias. The values of $\mathbf{w}$ and $b$ are not known and they need to be determined by training the network. The goal of the training process is to minimise the  mean squared error (MSE)

\begin{eqnarray}\label{eq:MSE}
MSE=\frac{1}{N} \sum_{i=1}^N \left( Y_i-\hat{Y}_i\right)^2,
\end{eqnarray}

\noindent where $N$ is the number of observations used to train the  network, $\hat{Y}_i$ is the prediction of the network for observation $i$ while $Y_i$ is the actual value for observation $i$. Notice that $Y$ can be the result of an experiment or of a theoretical calculation.

A feed-forward single-layer neural networks with a non-polynomial activating function  can approximate \emph{any} function \cite{LESHNO1993861}: the quality of the approximation will depend on the number of neurons, but also to some extent on the adopted training procedure~\cite{optimizers} or even the initialisation for weights and biases \cite{init}. The theorem given in Ref.\cite{LESHNO1993861} is independent on the details of the training procedure. However, the time required to train a large neural network can be prohibitive and the scarcity of data can limit the applications of NN. It is thus important to train the NN using the best \emph{features} in order to maximise the quantity of information one can extract from the data. See discussion in  Ref.~\cite{carnini2020trees} for more details.

To better illustrate such a concept, we present a simple toy-model. The calculations were performed using a Jupiter notebook that is provided as a Supplementary Material. The application of NN to the nuclear case starts at Sec.\ref{sec:nuclear}.

\subsection{Fitting a parabola}\label{sec:fit:para}

We generate 100 points in the interval $x\in[0,1]$ and evaluate the function $Y(x)=x^2$ . We split the data set into a \emph{training} and \emph{validation} as 80\% and 20\% of the data.
For the purpose of this example, we build a single layer NN using 8 neurons. We selected the  rectified linear  (ReLu) $f(x)=max(0,x)$~\cite{karlik2011performance,dorogush2018} as activation function. This is a quite popular choice since it allows to train networks with several layers without incurring in gradient vanishing problems.
The weights are initialised using a \emph{glorot uniform}~\cite{glorot2010understanding}, \emph{i.e.} they are drawn from a truncated uniform distribution. This is the default option for the Dense function in Keras~\cite{keras}.
Different initialiser could be used with more stringent assumptions on the structure of the weights, here we use a simple Bayesian approach: since we have no prior knowledge of the weights, we use a simple uniform distribution.

\begin{figure}
\begin{center}
\includegraphics[width=0.4\textwidth,angle=0]{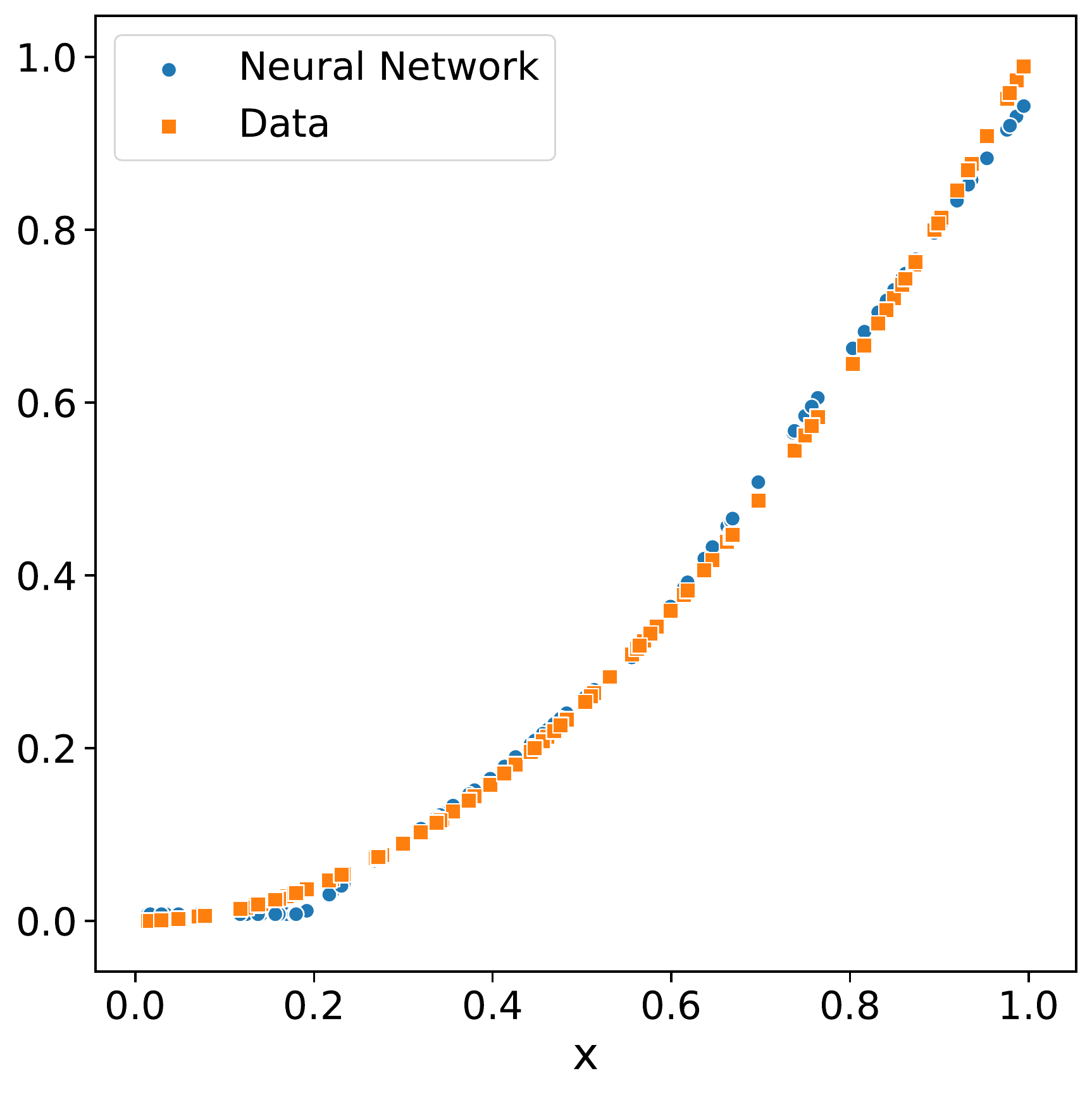}
\includegraphics[width=0.4\textwidth,angle=0]{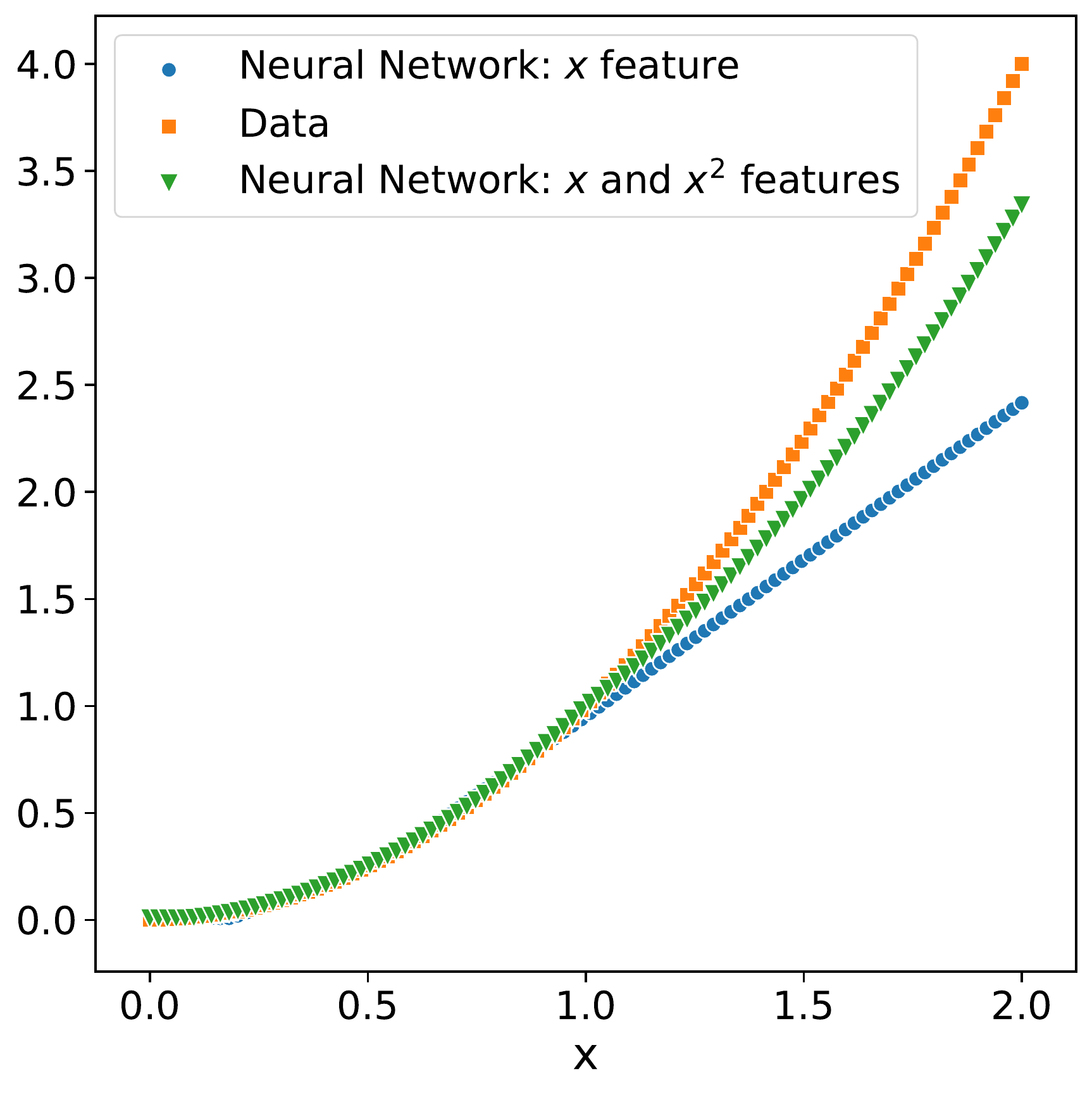}
\end{center}
\caption{Left: comparison between Neural Network interpolation (circles) and the original data points of the model (squares). Right: comparison between Neural Network extrapolations with (lower triangles) and without additional features (circles) and data points extracted from the original function $Y(x)=x^2$ (squares). See text for details.}
\label{parabola}
\end{figure}

After training the NN over 2000 epochs, we obtain the result given in the left panel of Fig.\ref{parabola}. We observe that the network is capable to grasp the structure of the data. 
The mean square error, Eq.\ref{eq:MSE}, on the training set is  $\approx5\cdot10^{-5}$. For this particular case, the MSE on the validation is also very similar showing that the NN provides a very good approximation of the data.

However, it is worth investigating what happens when we extrapolate using the NN in a region of space where there are no training data.
In the right panel of Fig.\ref{parabola}, we illustrate the extrapolation of the NN in the interval $x\in[1,2]$ and we compare it with the \emph{true} model $Y(x)=x^2$.
We see that the extrapolation quickly deteriorates and eventually the difference between \emph{true} model and NN quickly increases.
In particular, we notice that the NN is not able to capture the quadratic behavior of the data and we see a clear linear dependence in the region of the extrapolation, a side effect of choosing ReLU as an activation function.

 NN can \emph{learn} any type of structure in the data~\cite{LESHNO1993861}, but we can help the network by providing extra information, exactly in the same way as we did in Ref.~\cite{carnini2020trees}. In data science lexicon, this is called \emph{feature engineering}.
 The role of feature engineering is to improve the predictions obtained with the NN without increasing the number of data or changing its architecture.
 
 For this example, we train a new NN using exactly the same layout, but now including as input data both $x$ and $x^2$. In this case, we know the \emph{exact} structure of the toy model, but in a realistic case one should explore various possibilities. 
 
We stress that we are not changing the data, but we are simply making a transformation to highlight possible patterns in the training set. In other terms, we are only adapting the data \emph{representation} given to the algorithm (NN) that we chose. For example, including $x^2$ would not alter significantly the performance of a Decision Tree ~\cite{carnini2020trees}.

In the right panel of Fig.\ref{parabola}, we compare the trained network with the additional feature $x^2$ (triangles) to the simpler NN (full circles).
In the region $x\in[0,1]$ both NN do very well, but the one using additional features clearly behaves way better in the interval $x\in(1,2]$.

Finding the most relevant feature to improve the quality of the network during the training process is not an easy task. It is thus important to assess the quality of the NN, by defining error bars that may guide us in evaluating the quality of an extrapolation in a realistic case, where we do not know the structure of the \emph{exact} model. Finally, it is worth mentioning that adding \emph{features} that are not relevant for the model could actually lead to a deterioration of the results. See the discussion in Ref.\cite{mlmastery}.

\subsection{Error bars}

Within scientific literature there is no consensus on how to estimate error bars for NN. The standard approach based on the covariance matrix~\cite{dobaczewski2014error} can not be applied due to the typical large number of parameters and the clear difficulties in performing numerical derivatives in parameter space~\cite{roca2015covariance,shelley2019advanced}. In the following, we investigate three possible methods using the example illustrated in Sec.\ref{sec:fit:para}. For simplicity we  consider $x$ as the only feature to train the network.

\subsubsection{Epoch averaging}\label{epoch}

During the training of a NN, it is possible to store its coefficients at fixed values of epochs during the training~\cite{betterdeeplearning}. The approach was introduced independently by Polyak~\cite{polyak, polyak2} and Ruppert~\cite{ruppert}, and in also known as Polyak or Polyak-Ruppert averaging. 
In the case of this toy model, there is no variability, since the coefficients converge to their final result after few epochs and the gradient vanishes (see the Supplemental material). In a \emph{realistic} case, as shown in Fig.\ref{evo}, one observes that the MSE as a function of epochs decreases till reaching a \emph{plateau} where it starts fluctuating since the gradient does not reach a stable minimum. This means that the coefficients of the network are still slightly varying as a function of the epochs.
By storing them, for example, every 1000 epochs, we effectively create different networks with similar MSE. 
Having access to the different networks, we define the error bar in a statistical way as the interval where 68\% of predictions lie.

The main advantage of this method is that we do not need to train additional networks and thus it is a remarkable gain in CPU/GPU time. The downside is that all these networks are not independent from one another, but they typically manifest a strong degree of correlation and thus leading to an underestimation of the error bars. The additional downside is that if the gradient vanishes during the training, the NN gets to a stable configuration and as such averaging over successive epochs does not introduce any variability. This would also lead to an underestimation of the error bars.

\subsubsection{Bootstrap}

A very simple alternative to the epoch averaging is based on bootstrap~\cite{efron1986bootstrap,pastore2019introduction}. Since the training of a network is essentially a non-linear fit, we can explore the landscape of parameter space by using slightly different training sets. To this purpose, given the data, we create 100 sets of validation/training sets, by random sampling the original one.
For each of them, we train a NN with the same architecture. We see that each individual network will be trained only on a fraction of the total data (here 80\%), but the full ensemble will be trained over \emph{all} the data.
As a consequence, using bootstrap, we maximise the information contained in the data. Having access to 100 networks, we average them out and define an error bar as the region where 68\% of the curves lie, while the expected prediction is obtained by simply averaging out the outcomes of all NNs.

\begin{figure}
\begin{center}
\includegraphics[width=0.4\textwidth,angle=0]{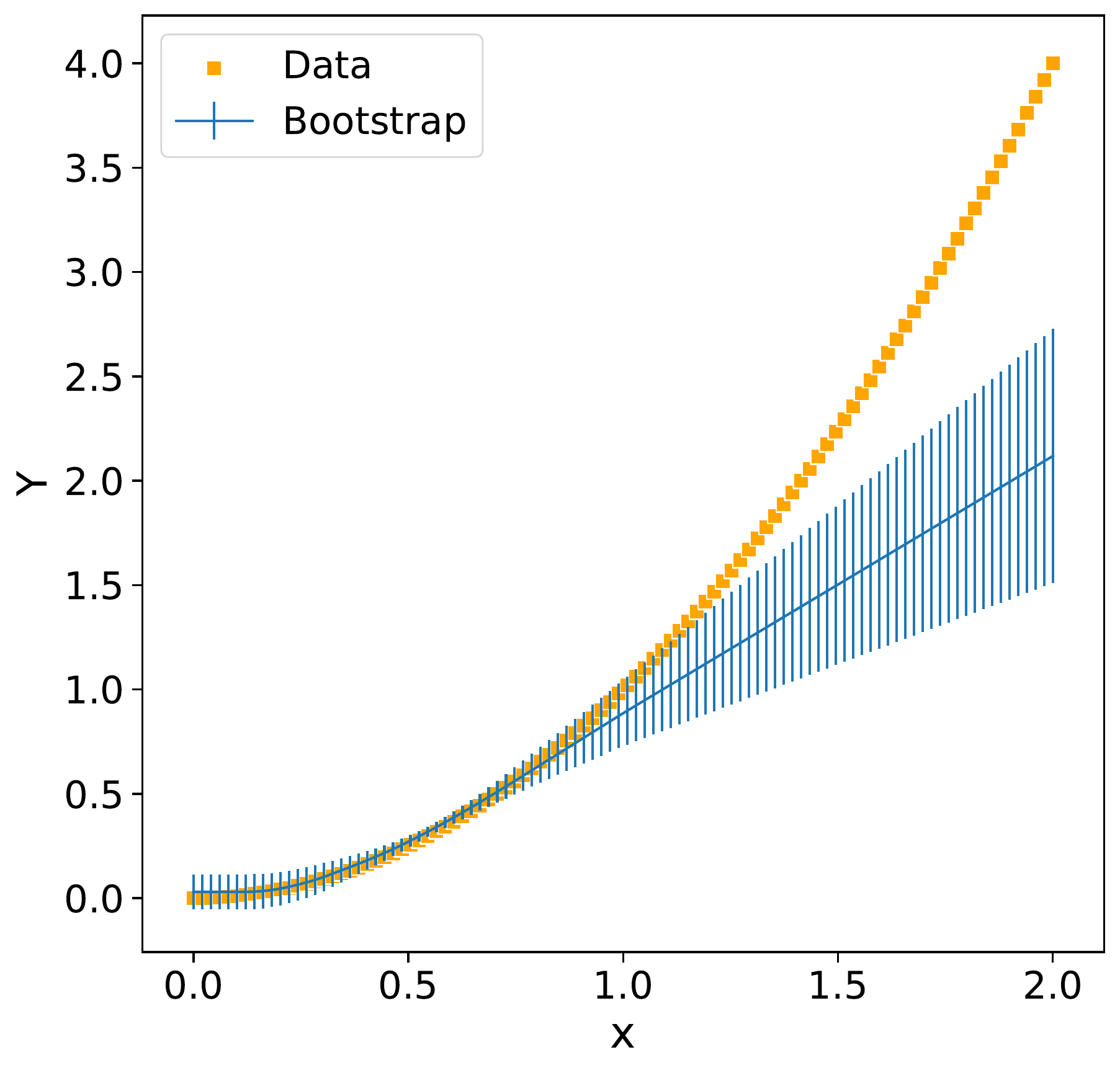}
\includegraphics[width=0.4\textwidth,angle=0]{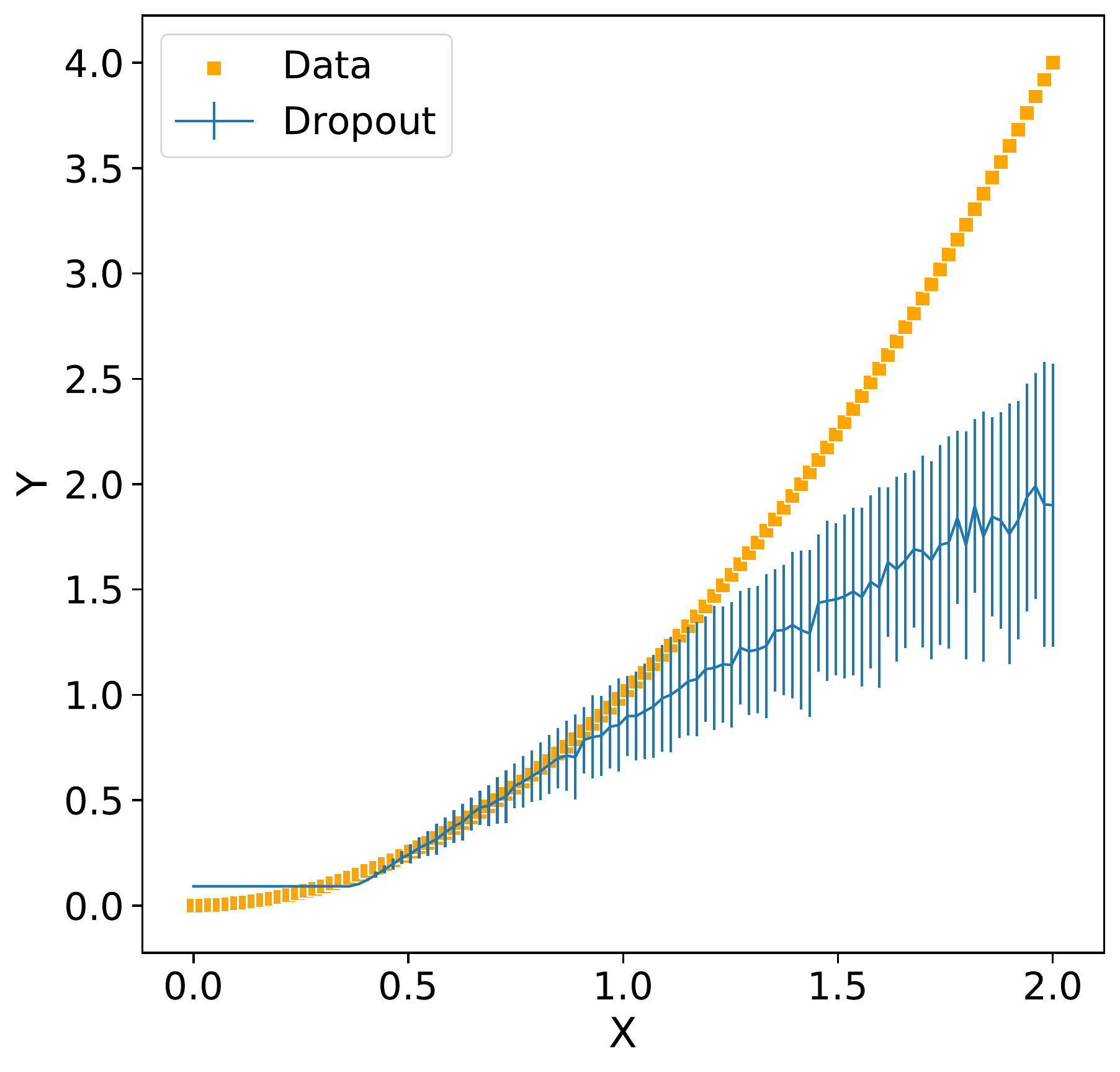}
\end{center}
\caption{Left: comparison between Neural Network using bootstrap (solid line) and the original data (squares). Right: comparison between Neural Network using dropout  (solid line) and the original data (squares). See text for details.}
\label{boot_drop:parabola}
\end{figure}

In the left plot of Fig.\ref{boot_drop:parabola}, we show the resulting prediction obtained using the bootstrap method: we see that the error bars are very small in the region $[0,1]$ while they grow when $x$ approaches the limits of the data set.
Beyond $x=1$, the error bars grow remarkably fast, as expected due to the lack of information in this region of space.

The error bars obtained with bootstrap strongly depend on the variability of the predictions. By investigating in more detail the left plot of Fig.\ref{boot_drop:parabola}, we observe that the \emph{true} model falls within the error bars only up to $x\approx1.25$. To go beyond this point one should use larger error bars by taking, for example, two standard deviations, but one clearly notices that this makes the prediction less and less relevant given the size of the error bars.

\subsubsection{Dropout}

Finally, we consider the dropout method to estimate the error bars.
This is a technique that has been introduced to avoid over-fitting and to improve predictions. The idea is very simple~\cite{srivastava2014dropout}: we train the network over the training set and we randomly switch off some neurons. In this simple toy model we turn off 1 neuron each time.

We call the trained NN 100 times allowing to switch off one neuron randomly each time. In this way we produce 100 predictions and we define the error bar, as in the bootstrap case, as the region where 68\% of the predictions lie.
The result is illustrated in the right plot of Fig.\ref{boot_drop:parabola}.

By comparing the results for bootstrap and dropout in Fig.\ref{boot_drop:parabola},  we see that the dropout gives very similar results to the bootstrap case. The error bars estimated in this way contain the \emph{true} model up to $x\approx1.25$. Beyond this point although the error bars continue growing, the difference between the model and the prediction is clearly under-estimated by our error bars.

We now move to some realistic nuclear data to continue testing the three methods presented here to evaluate error bars.

\section{Nuclear masses}\label{sec:nuclear}

Currently~\cite{wang2017ame2016}, more than 2400 nuclear masses have been experimentally measured with very high degree of accuracy. The exact knowledge of nuclear binding energies play a crucial role in several physical scenarios as for example r-process nucleosynthesis~\cite{cowan1991r} or in the determination of the chemical composition of the crust of a neutron star~\cite{chamel2008physics}. Since NN have been recently applied to perform extrapolations of nuclear masses~\cite{uta16,uta17,neufcourt2018bayesian,pastore2020impact} in regions where no experimental data are still available, we find very important to provide a reliable estimate of the error bars to help evaluating the quality of such results.

As done in the previous section, we  validate the quality of the extrapolation against a \emph{closed-form} model. To this purpose, instead of using experimental masses, we use binding energies extracted from a nuclear model.
The synthetic data are generated using a liquid-drop model~\cite{pastore2019introduction} and the NEDF via the  Skyrme SLy4 parametrisation~\cite{chabanat1998skyrme,stoitsov2006large}. 
Both LD and SLy4 give a reasonable reproduction of nuclear binding energies, with an RMS of few MeV.
Other mass-models with improved accuracy are available within the literature~\cite{goriely2009first,duflo1995microscopic,moeller19941992,sobiczewski2014predictive}, but -- for the present calculation -- we are only interested in considering two categories of models: one linear and the other non-linear, just to check if the performances of the NN are impacted by such a choice.

Both LD and SLy4 predict the existence of way more nuclei than the one experimentally observed~\cite{wang2017ame2016} and they allow us to perform benchmark against NN predictions along several complete isotopic chains.

To be as realistic as possible, we define for the training set of the NN all the measured isotopes given in Ref~\cite{wang2012ame2012}, but using the values of binding energies calculated by the models. We stress that at this stage, we are not interested in reproducing as accurately as possible experimental data, but to validate the approach for extrapolations based on NN.

We build a NN formed by 3 layers having 16-8-16 neurons respectively densely connected and using a ReLu activation function. No particular effort was dedicated to fine tune and optimise the architecture, except for fixing reasonable defaults (see for example~\cite{dorogush2018}). 
Following Ref.~\cite{anil2020nuclear}, the NN is directly trained on total binding energies per particle to avoid any additional bias introduced by the model itself.

\subsection{Liquid Drop data}

The LD model express the nuclear binding energy,$B$, as a sum of five different terms depending uniquely on proton ($Z$) and neutron ($N$) number as

\begin{eqnarray}\label{bene}
\frac{B^{LD}(N,Z)}{A}=a_v -a_sA^{-1/3}-a_c\frac{Z(Z-1)}{A^{4/3}}-a_a\frac{(N-Z)^2}{A^2}-\delta\frac{mod(Z,2)+mod(N,2)-1}{A^{3/2}}
\end{eqnarray}

\noindent where $A=N+Z$ and the coefficients $a_v,a_s,\dots$ have been adjusted in Ref.~\cite{pastore2019introduction}.
The features of the model are quite simple and one could use them to directly train the network~\cite{carnini2020trees}. However, here we consider no \emph{a priori} knowledge of the model and we use only $N$ and $Z$ as features. 

\begin{figure}
\begin{center}
\includegraphics[width=0.6\textwidth,angle=0]{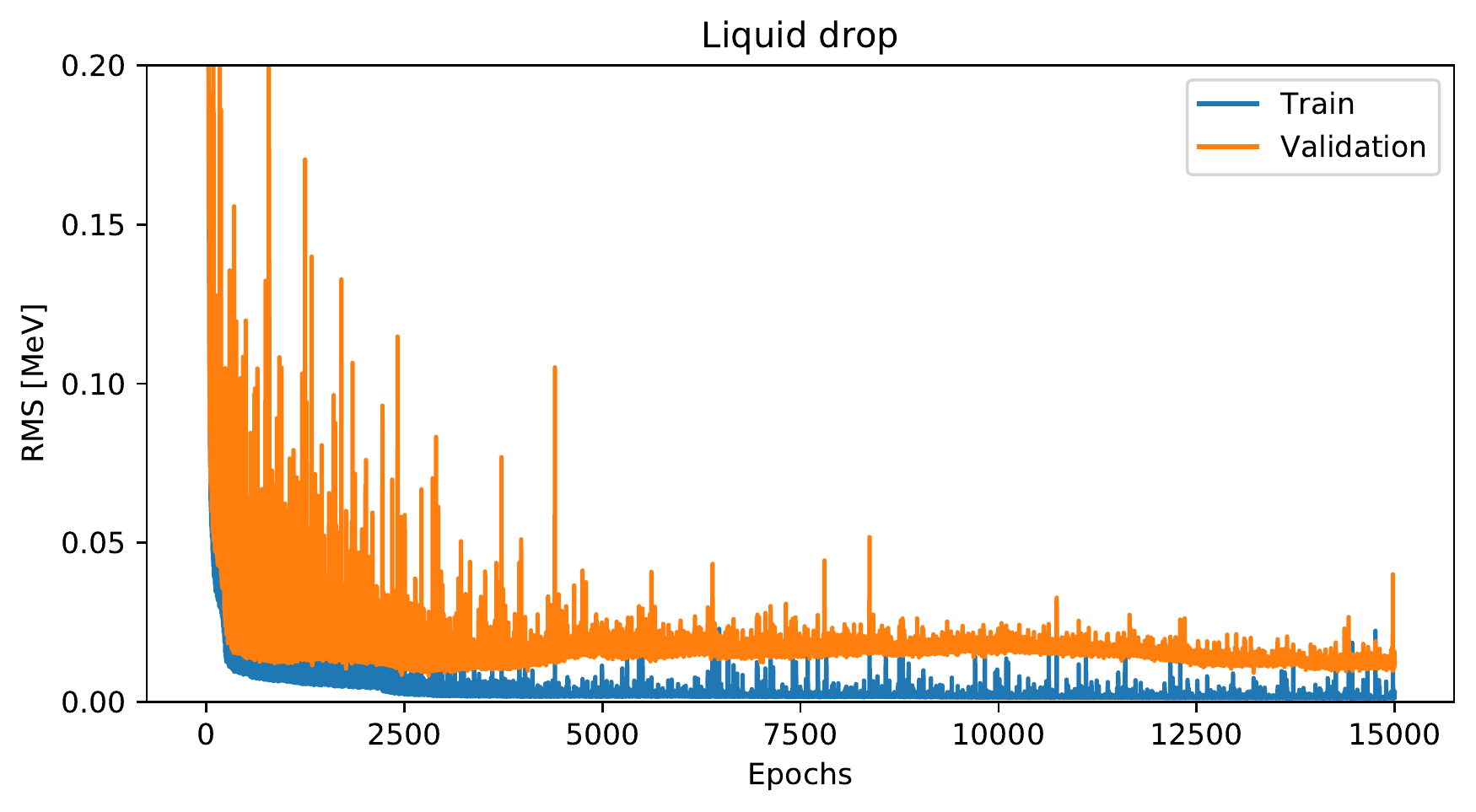}
\end{center}
\caption{Evolution of the root mean square, Eq.\ref{eq:MSE}, as a function of the epochs used for training the NN and based on LD data. See text for details.}
\label{evo}
\end{figure}

In Fig.\ref{evo}, we show the evolution of the RMS (expressed in MeV) as a function of the epochs, for both training and validation sets. We used as a label the energy per particle $B/A$ given by Eq.\ref{bene}. The motivation for our simple choice of the network can be seen here: after few thousands of epochs, the network has reached a \emph{plateau}, where the gradient is small.
The fast convergence is a necessary ingredient for our successive analysis on error bars.

The final RMS on the training set is $\sigma_{tr}\approx50$ keV, while on the validation is $\sigma_{val}\approx100$ keV. The accuracy is roughly of the same order of magnitude of the LD respect to experimental nuclear data. By training a second NN on the residuals it would be probably possible to further reduce the RMS~\cite{anil2020nuclear}, but this is not the goal of the present discussion. 
We want to stress here that the NN trained here is probably not the best one we can build out of the data, but a reasonable tool that we can use for our analysis on error bars.

\begin{figure}
\begin{center}
\includegraphics[width=0.43\textwidth,angle=0]{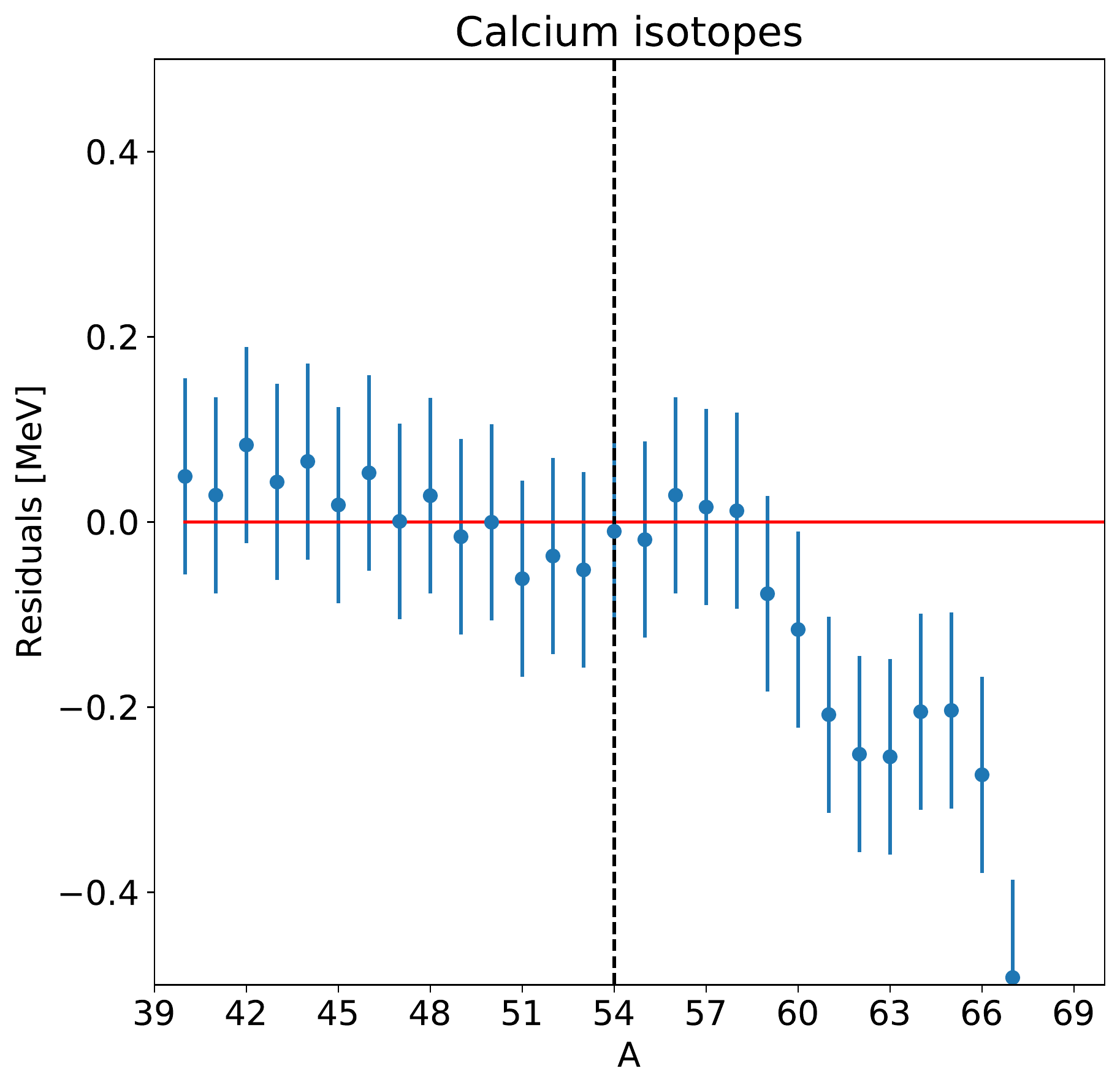}
\includegraphics[width=0.43\textwidth,angle=0]{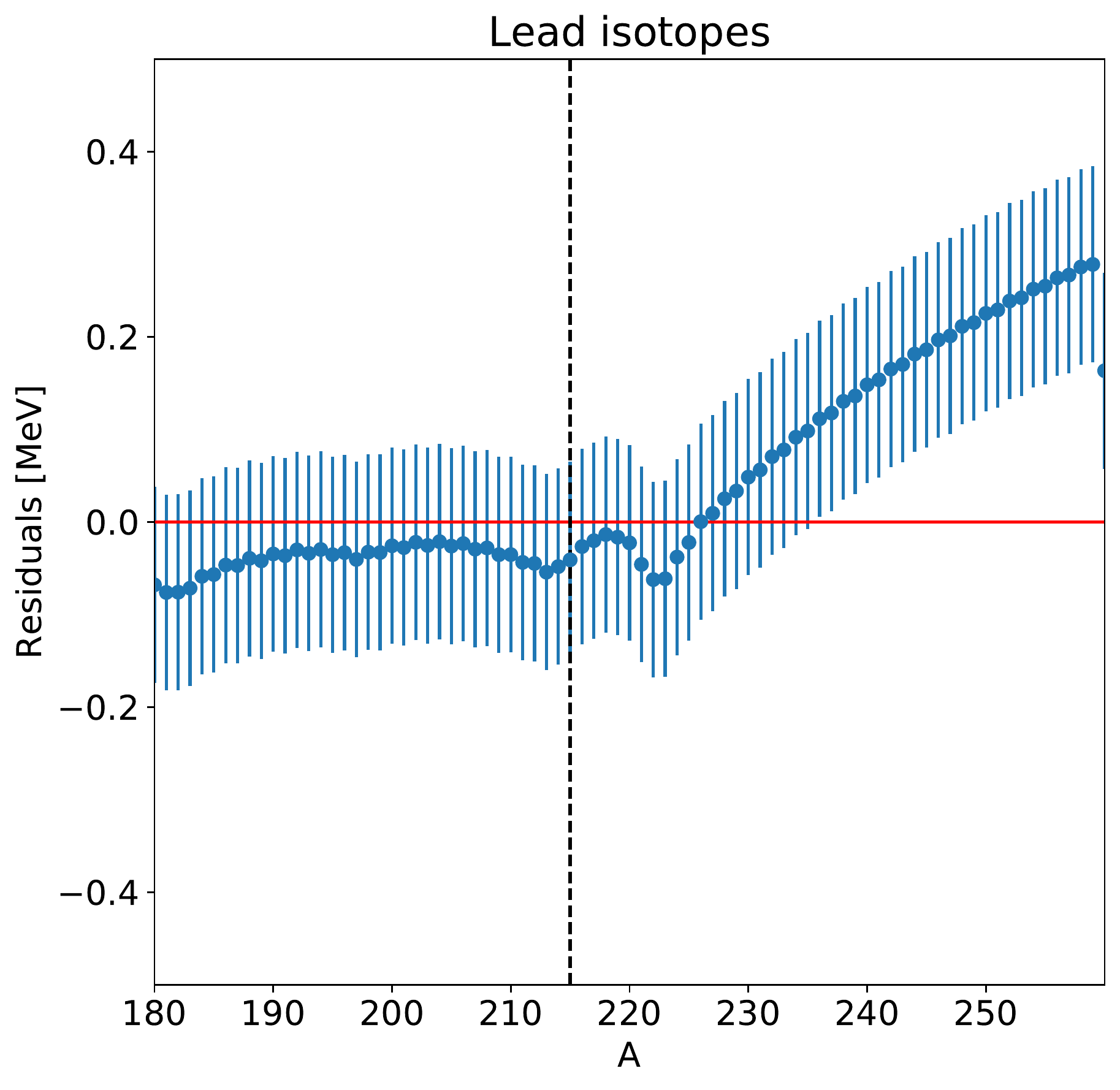}
\end{center}
\caption{Evolution of the difference between the energy per particle calculated using the LD model and the NN. The vertical dashed line indicates the position of the last  isotope used to train the NN. The horizontal line indicates the zero and it helps guiding the eye. See text for details.}
\label{residue}
\end{figure}

Having trained the NN, we define the residuals as the difference between the $B^{LD}/A$ and the binding energy per particle as calculated via the NN $B^{NN}/A$.
In Fig.\ref{residue}, we compare the evolution of the residuals as a function of the mass number $A$ for two isotopic chains: Calcium and Lead. 
The choice of the isotopic chains is somehow arbitrary: we picked those to illustrate that there is no difference when selecting a light or an heavy element. 

To guide the eye, we have added an horizontal line to indicate the position of zero. The vertical dashed line indicates the position of the heaviest isotope used to train the network.
The first estimate of the error of the NN is represented by its RMS. Assuming that the errors are normal, we set the error bar~\cite{barlow1989statistics} to 1$\sigma$, equal to the global RMS of the model on the validation set, as done in Ref.~\cite{pastore2020impact}.

The simple error bar used here can be considered probably as a very good approximation to compare with data within the range of the training, \emph{i.e.} in this case for Calcium isotopes with $A\le54$ and $A\le215$ for lead isotopes, while it clearly underestimates the real statistical error in the extrapolation region. The result shown in Fig.\ref{residue}, especially in the extrapolated region, is very much dependent on several quantities as the initialisation of the weights or a different splitting of the data in training/validation.
As such, the \emph{na\"ive} extrapolation done here can not be considered as reliable even if accidentally some of the nuclei in the extrapolated region are still well reproduced.

A more refined error bar, based on the methodologies of Sec.\ref{sec:fit:para} is presented in Sec.\ref{sec:error}.


\subsection{Skyrme data}

Within the NEDF theory, the total binding energy of a nucleus is written as the space integral of an energy density functional obtained using a microscopic Skyrme interaction~\cite{perlinska2004local,becker2017solution}. Differently from LD, the Skyrme model is non linear in parameter space,  since all the densities appearing in the various terms of the functional~\cite{tondeur1983skyrme} are obtained as a self-consistent solution of the Hartree-Fock-Bogoliubov equations using an iterative procedure~\cite{ring2004nuclear}. 
In the present article, we consider the data calculated in Ref.~\cite{stoitsov2006large} using the SLy4 functional~\cite{chabanat1998skyrme}.

\begin{figure}
\begin{center}
\includegraphics[width=0.6\textwidth,angle=0]{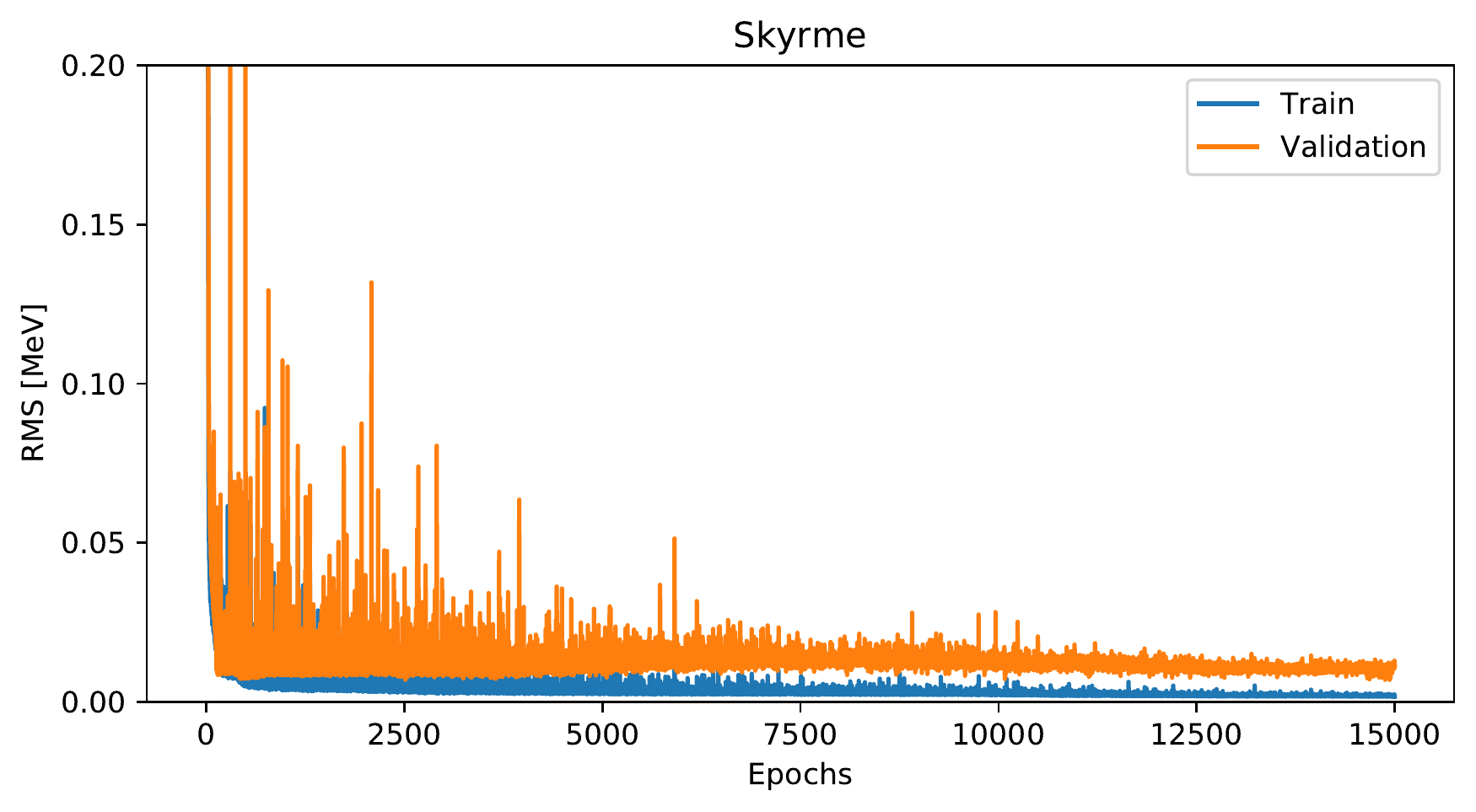}
\end{center}
\caption{Same as Fig.\ref{evo}, but using SLy4 data.}
\label{evo2}
\end{figure}

From the statistical point of view, the interest in using a microscopic calculation is related to the fact that the \emph{features} are not so evident as in the case of the LD model shown in Eq.\ref{bene}, although the overall quality is reproducing nuclear masses is similar.

Following the same procedure used for LD, we now train a NN over Skyrme data with the same layout. In Fig.\ref{evo2}, we illustrate the evolution of the RMS as a function of the number of epochs used for the training. 
As discussed previously, the goal of the current paper is not to find the optimal NN, but to discuss the optimal methodology to estimate error bars.

After 15000 epochs, we obtain an RMS of $\sigma_{tr}=\approx50$ keV on the energy per particle for the training  and $\sigma_{val}=\approx100$ keV for the validation set. These performances are comparable to the one of the NN trained on the LD data.

\begin{figure}
\begin{center}
\includegraphics[width=0.43\textwidth,angle=0]{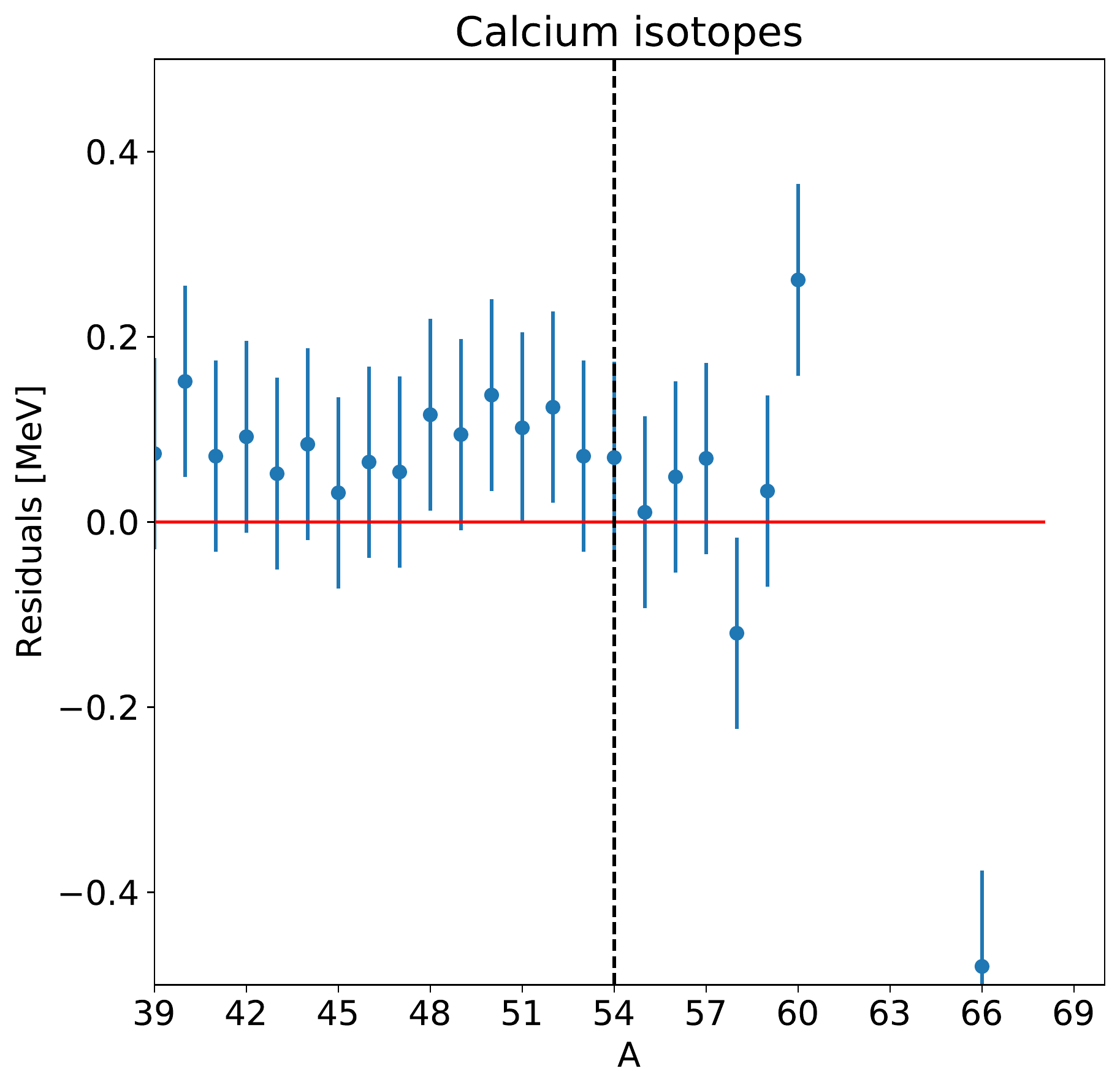}
\includegraphics[width=0.43\textwidth,angle=0]{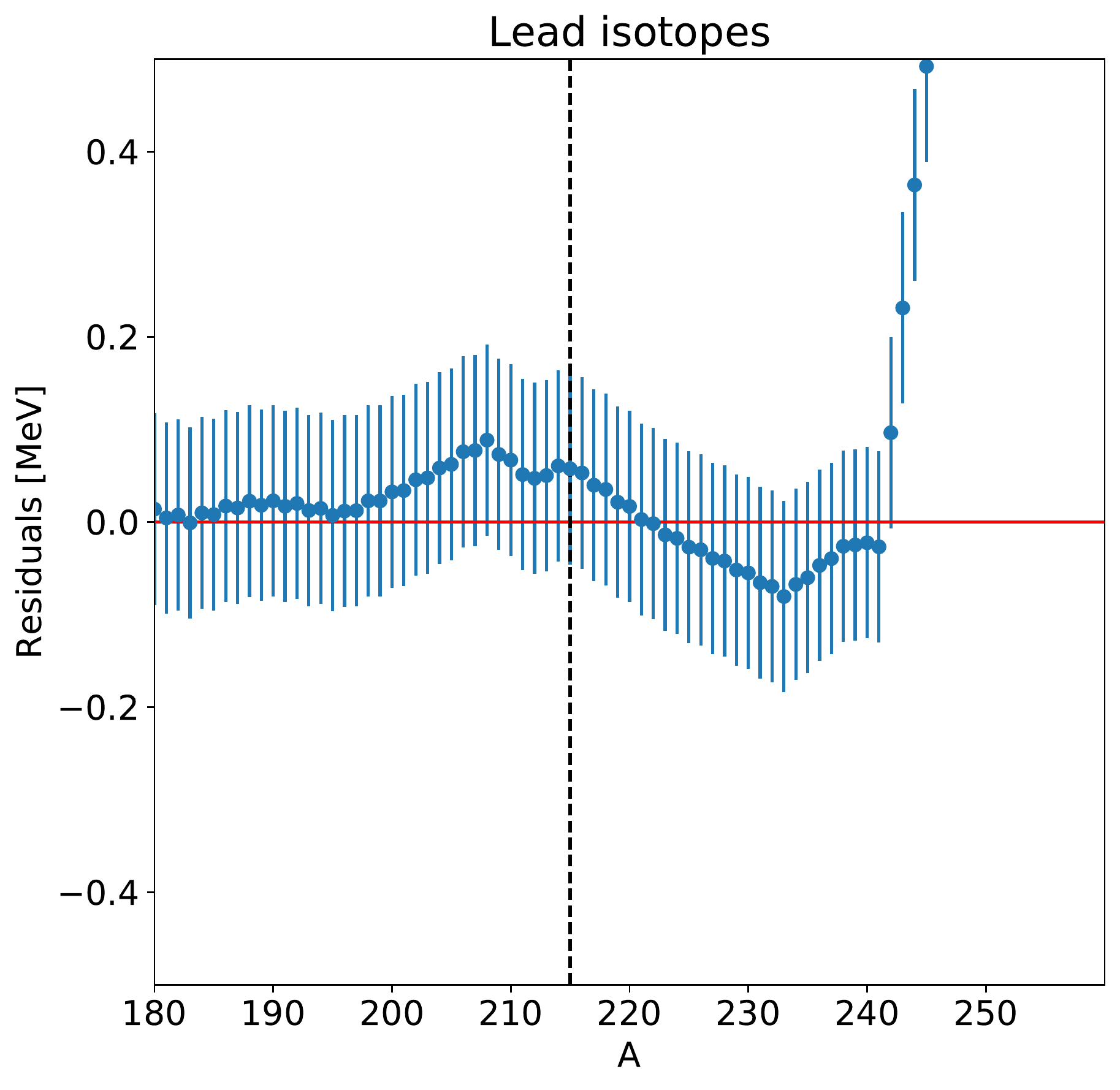}

\end{center}
\caption{Same as Fig.\ref{residue}, but for SLy4 data}
\label{residuesly4}
\end{figure}

In Fig.\ref{residuesly4}, we show the evolution of the NN predictions along the isotopic chain of Calcium and Lead, in the same way as we did in Fig.\ref{residue}.
Although the drip-lines obtained using LD and SLy4 are not equal, we observe that our NN behaves quite nicely for the first isotopes beyond the last known nucleus and then it diverges. 

The simple error estimate based on the RMS clearly under-estimate the true error, although by chance the Lead isotopes are very well reproduced by our NN in the extrapolated region.

\section{Error estimate}\label{sec:error}

In this section, we apply the three different methods discussed in Sec.\ref{sec:fit:para} to the realistic data obtained from LD and Skyrme-SLy4 models.

\subsection{Epoch-averaging}

By looking at Fig.\ref{evo} and Fig.\ref{evo2}, we observe that both loss functions reach a \emph{plateau} region around 10000 epochs. The gradient is not zero and we still observe small fluctuations.
This means that the weights and the biases of the NN are still fluctuating as a function of the epochs, although we expect these fluctuations to be small.

We take advantage of the epoch-averaging idea presented in Sec.\ref{epoch},by continue the training and store the weights every 1000 epochs for  100 times. We create 100 models with all slightly different parameters, but using exactly the same data set for training and validation.

We define an \emph{average} model, by calculating the mean value of the models and  the 1-$\sigma$ error as the region containing 68\% of the predictions. The result is presented in Fig.\ref{epoch} for Ca and Pb isotopic chains using the synthetic data calculated using LD  and Skyrme.
The vertical lines indicate the position of the last experimentally known nucleus in the chain.

\begin{figure}
\begin{center}
\includegraphics[width=0.43\textwidth,angle=0]{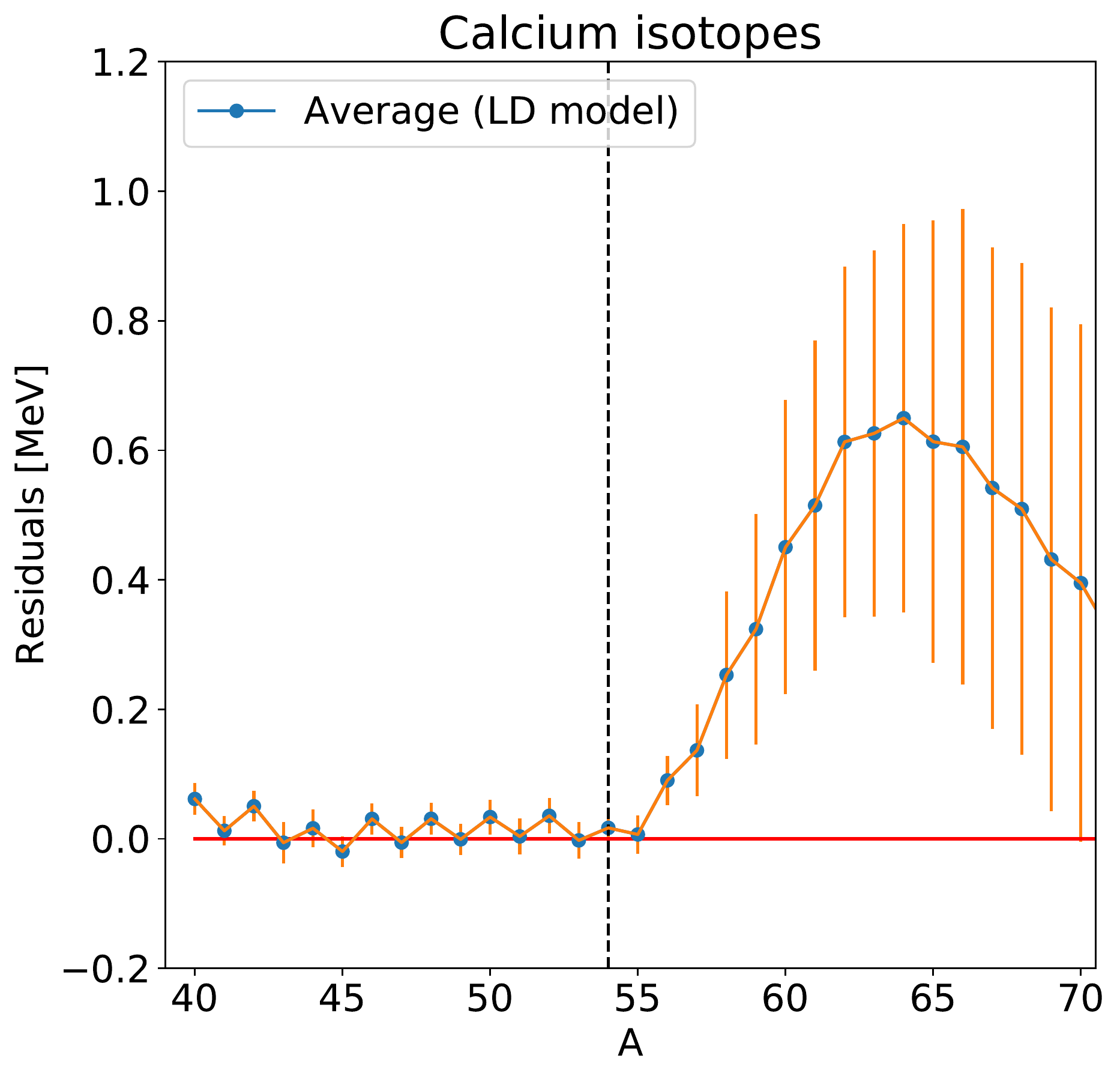}
\includegraphics[width=0.43\textwidth,angle=0]{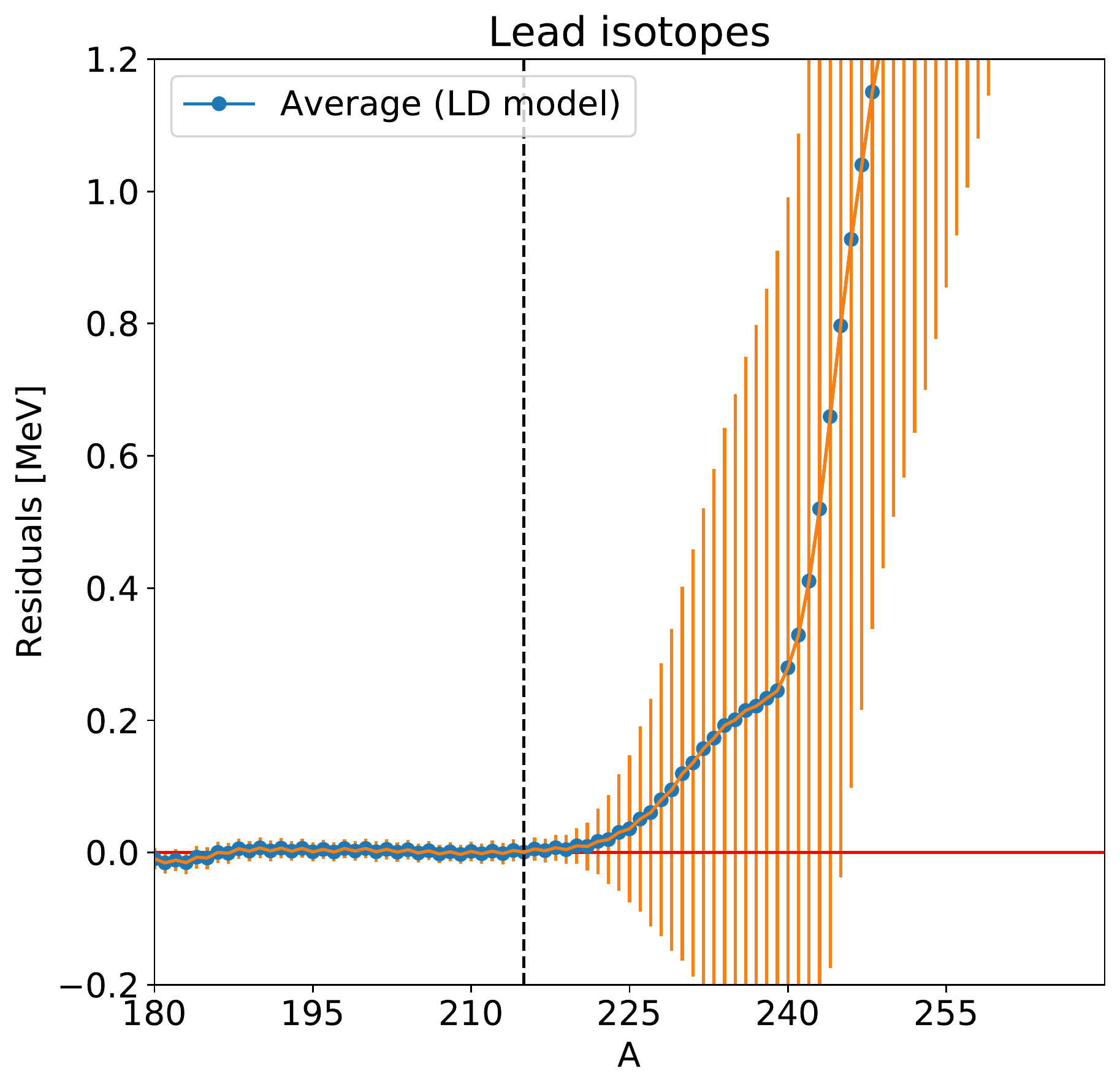}\\
\includegraphics[width=0.43\textwidth,angle=0]{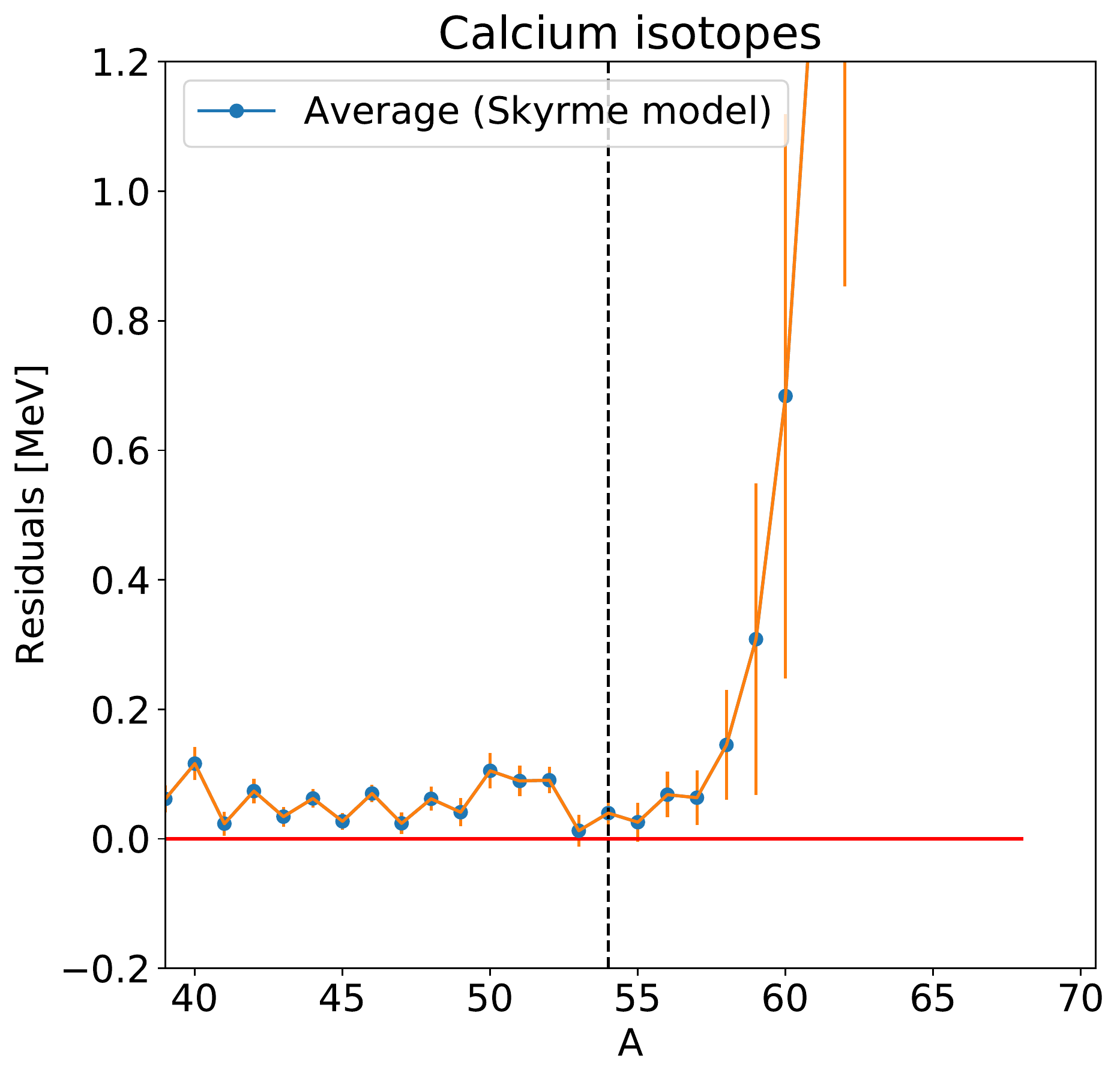}
\includegraphics[width=0.43\textwidth,angle=0]{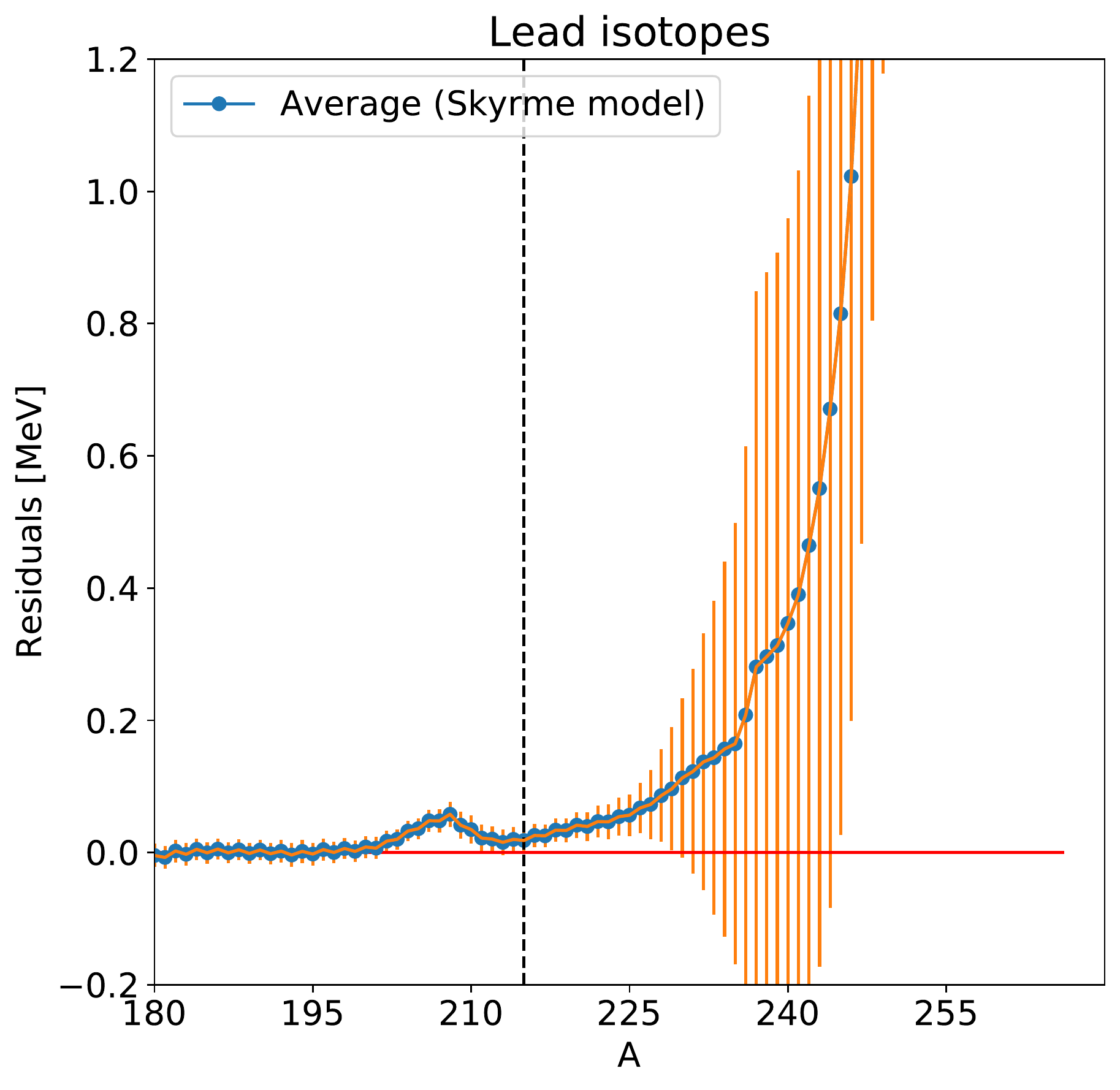}
\end{center}
\caption{Same as Fig.\ref{residue}, but using epoch-averaging to estimate error bars. On the left panels calcium isotopes and on the right panels the lead isotopes. The top row has been obtained using LD data, while the bottom row using Skyrme data. See text for details.}
\label{epoch}
\end{figure}

\noindent We observe that the error bar is very small in the region where data are present (the training set). This value is even smaller than the na\"ive RMS shown in Figs.\ref{residue}-\ref{residuesly4}. In the region of extrapolation, \emph{i.e} $A\ge54$ for calcium isotopes and $A\ge215$ for lead, the error bars  start to grow quite fast becoming soon way larger than the simple RMS.
The predictions done with the NN are \emph{reasonable} for isotopes with few neutrons beyond the vertical line, but the prediction quickly deteriorates and the uncertainties becomes soon very large, and in clear disagreement with the \emph{true} model.

\noindent Although such a procedure is quite simple and not too expensive in terms of CPU/GPU, it is worth recalling that the different networks are not independent from one another, but they are highly correlated. This can be checked by evaluating the correlation matrix between them. 
The consequence is that changing the training set will lead to a different prediction and different error bars. Although the main outcome will remain the same.

\subsection{Bootstrap}

A different approach to avoid the strong dependence on the training set is based on  the bootstrap method~\cite{efron1986bootstrap,pastore2019introduction}. In this case, we randomly split the available data into training and validation, reshuffling them at each bootstrap iteration.
By using 100 bootstrap iterations, we have thus obtained 100 neural networks, all trained for the same amount of epochs (15000).
By visually inspecting the evolution of the RMS, we have checked that all networks have reached convergence with a final RMS on the training and validation set of the same quality of the original one. As done before, we calculate the average and the variance of the various neural networks to define an error bar.
In this case, by examining the correlation matrix, one observes still a correlation, but much weaker than in the previous case of epoch averaging. The different NN are still partially correlated since they have been trained to strongly overlapping data-sets.

\begin{figure}
\begin{center}
\includegraphics[width=0.43\textwidth,angle=0]{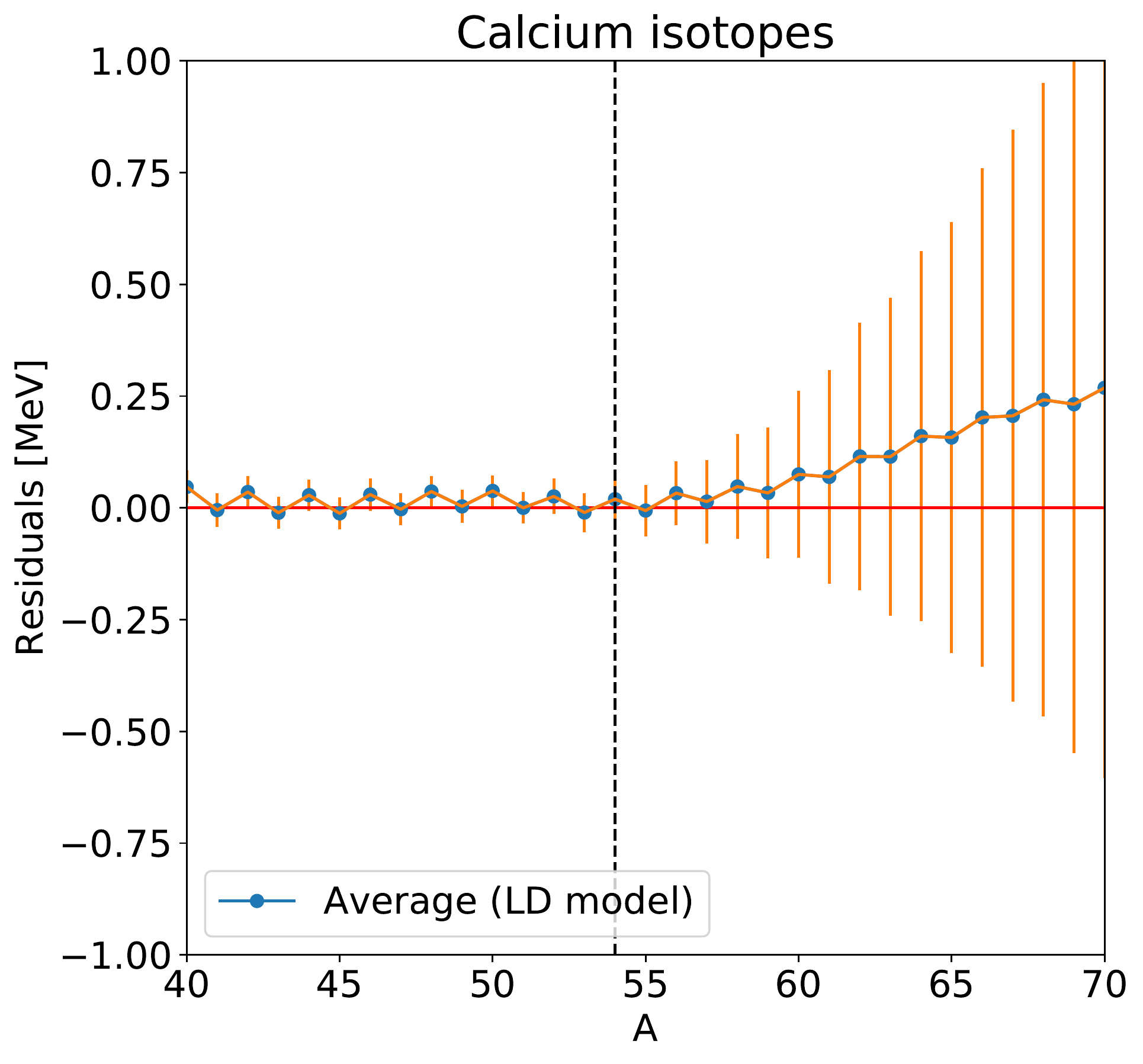}
\includegraphics[width=0.43\textwidth,angle=0]{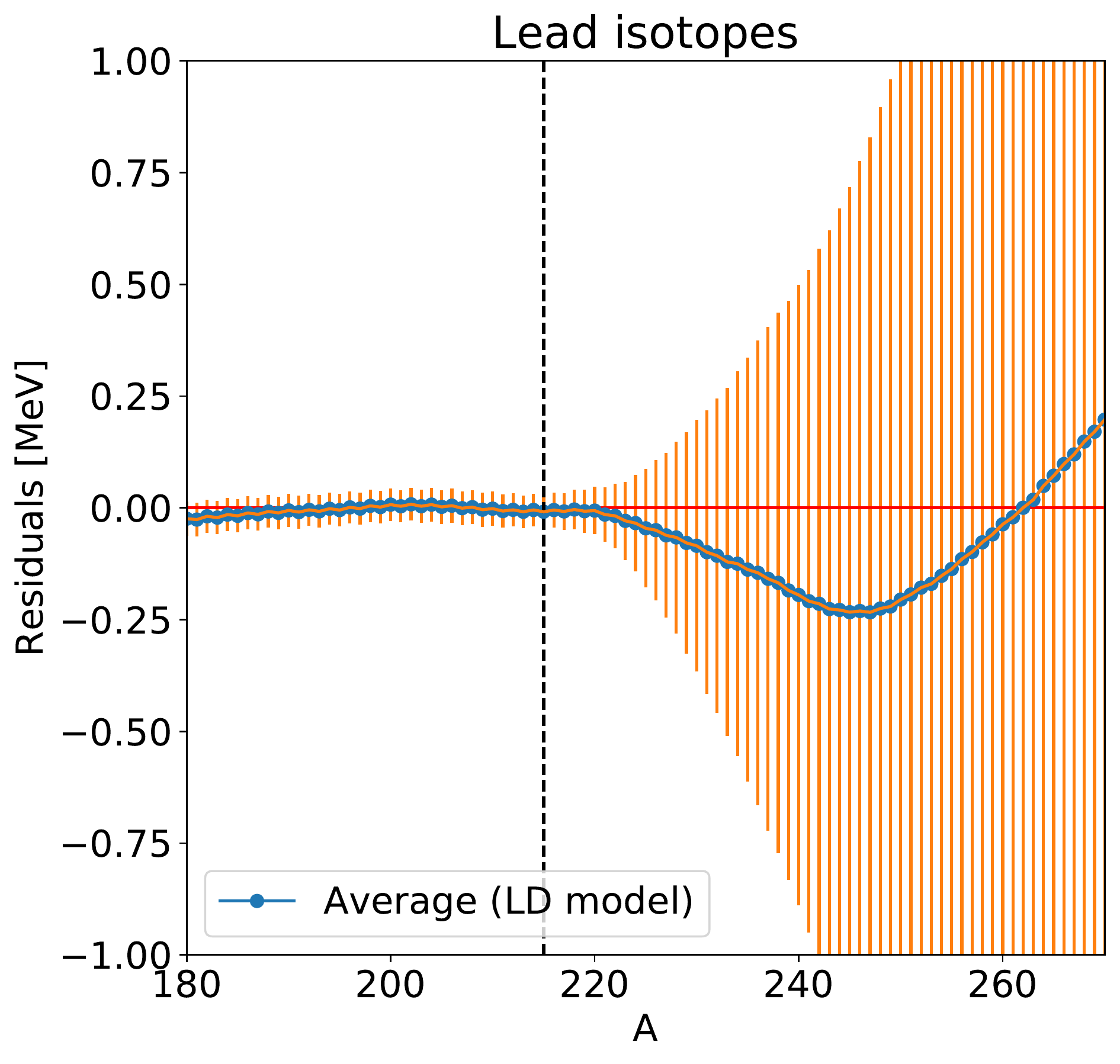}\\
\includegraphics[width=0.43\textwidth,angle=0]{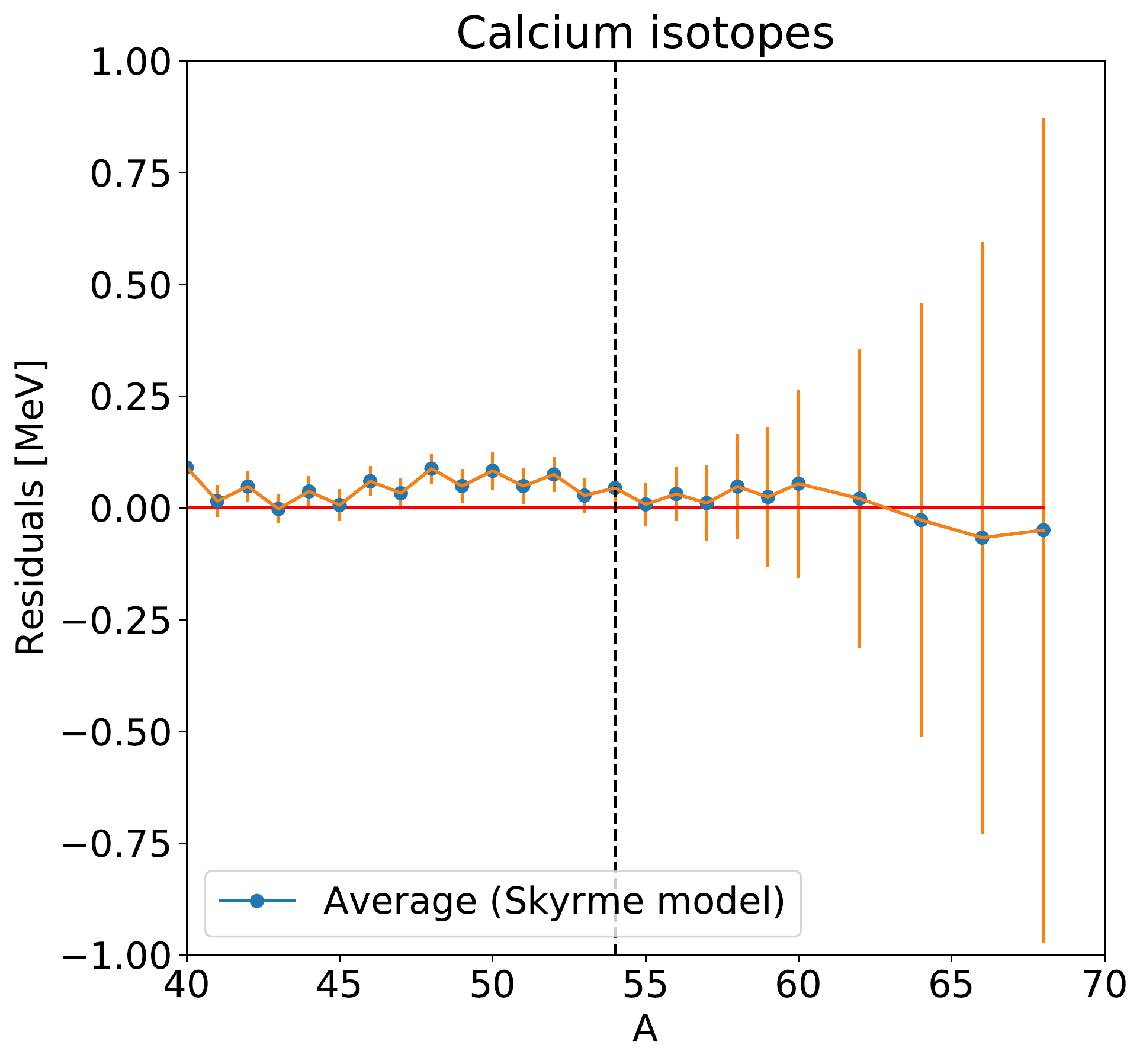}
\includegraphics[width=0.43\textwidth,angle=0]{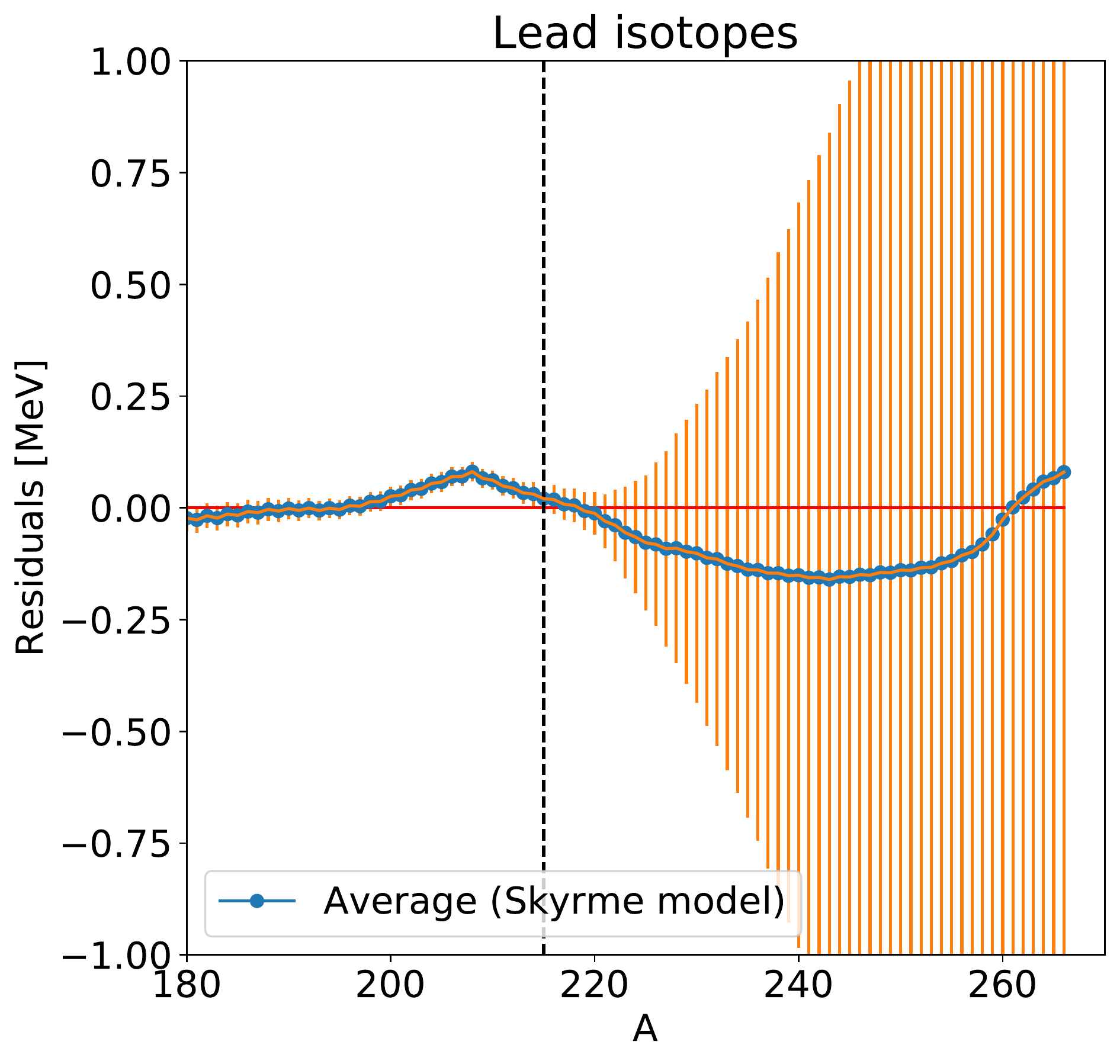}
\end{center}
\caption{Same as Fig.\ref{residue}, but using bootstrap to estimate error bars. See text for details.}
\label{boot}
\end{figure}

In Fig.\ref{boot}, we illustrate the evolution of the difference between the energy per particle as calculated with the LD and Skyrme models using the NN, together with the associated error bar trained using the bootstrap method.

The error bars obtained with bootstrap are robust, \emph{i.e.} they do not depend on the particular choice of the initial data-set, but they seem to overestimate the real precision of the NN. By moving few isotopes beyond $A=215$ for Pb and $A=54$ for Ca, the error grows fast and they soon reach several hundreds of keV. From the statistical point of view this error bar can be seen as a very conservative estimate of the predictive power of the model since the \emph{true} model is always included within these error bars. This was not the case of epoch-averaging as shown in Fig.\ref{epoch}.

\subsection{Dropout}

We now consider the third method to estimate error bars and based on dropout~\cite{srivastava2014dropout}. As explained in detail in the simple toy-model example,  during the training of the NN, we randomly turn off  a given percentage of neurons for the predictions. This procedure is used in the literature to check the robustness of the network and to avoid over-fitting of the weights. As discussed in Ref.~\cite{srivastava2014dropout}, dropout can be used as an approximation to a more involved Bayesian Neural Network (BNN). The main advantage of dropout compared to BNN is that the typical training time required to determine the weights is shorter.
To mimic BNN, we apply the same dropout also to the prediction. This means that every time we call the NN to obtain a value, we randomly turn off a set of neurons. This will give us the required variability to estimate an error. For simplicity (namely, writing the least amount of code), we used the option from Keras~\cite{keras, chollet2015keras} to use dropout at both training and prediction phases.

The layout of the NN is the same of previous cases, but we had to increase the number of epochs to 30000 to let the gradient get to a stable \emph{plateau}.  
The drop-out rate is arbitrary, for the present calculation we have used a rate of 5\%. Other rates could be explored, but given the architecture of the NN and the relatively small number of neurons, higher dropout would lead to very poor performances. In the present case, using 5\% dropout, the resulting root mean square is of $\approx150$ keV  for both the  training and the validation set. This value is roughly 3 times worst than what we obtained using the same layout without dropout.

During the training phase, some neurons are silenced, forcing the other neurons to learn to compensate the missing ones. By spreading the information learned by one neuron to another, the dropout effectively improves the performance by assembling the predictions of different models trained in parallel. By default, the dropout is not used during the predictions, and the output of all neurons is considered. In this way, the prediction process is deterministic and reproducible. 

If, however, the dropout is used at prediction time, every time a prediction is launched a different result is possible. This leads to a distribution function for each and every prediction. If we assume a normal distribution, by averaging and taking the standard deviation an estimation with a bar error can be obtained. Of course, the normal distribution is not strictly required, but in order to assess the root mean square $\sigma$, the second momentum should be finite. %

With a finite amount of neurons, the possible outcomes are limited: for a neural network with 10 neurons and a drop out rate of 10\%, every time a prediction is made one of the neurons is silenced. This lead to only ten possible distinct values for the prediction.

\begin{figure}
\begin{center}
\includegraphics[width=0.43\textwidth,angle=0]{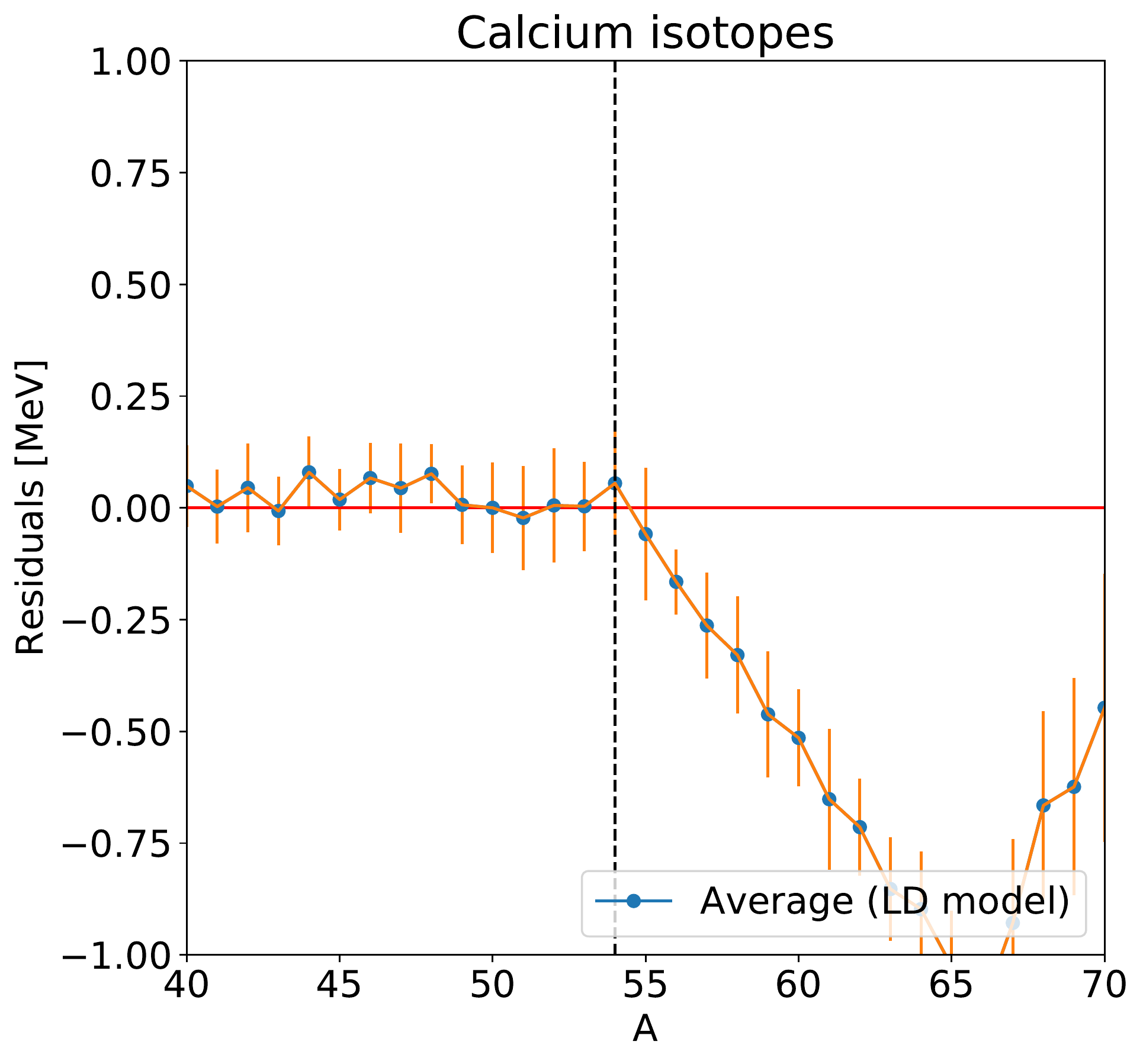}
\includegraphics[width=0.43\textwidth,angle=0]{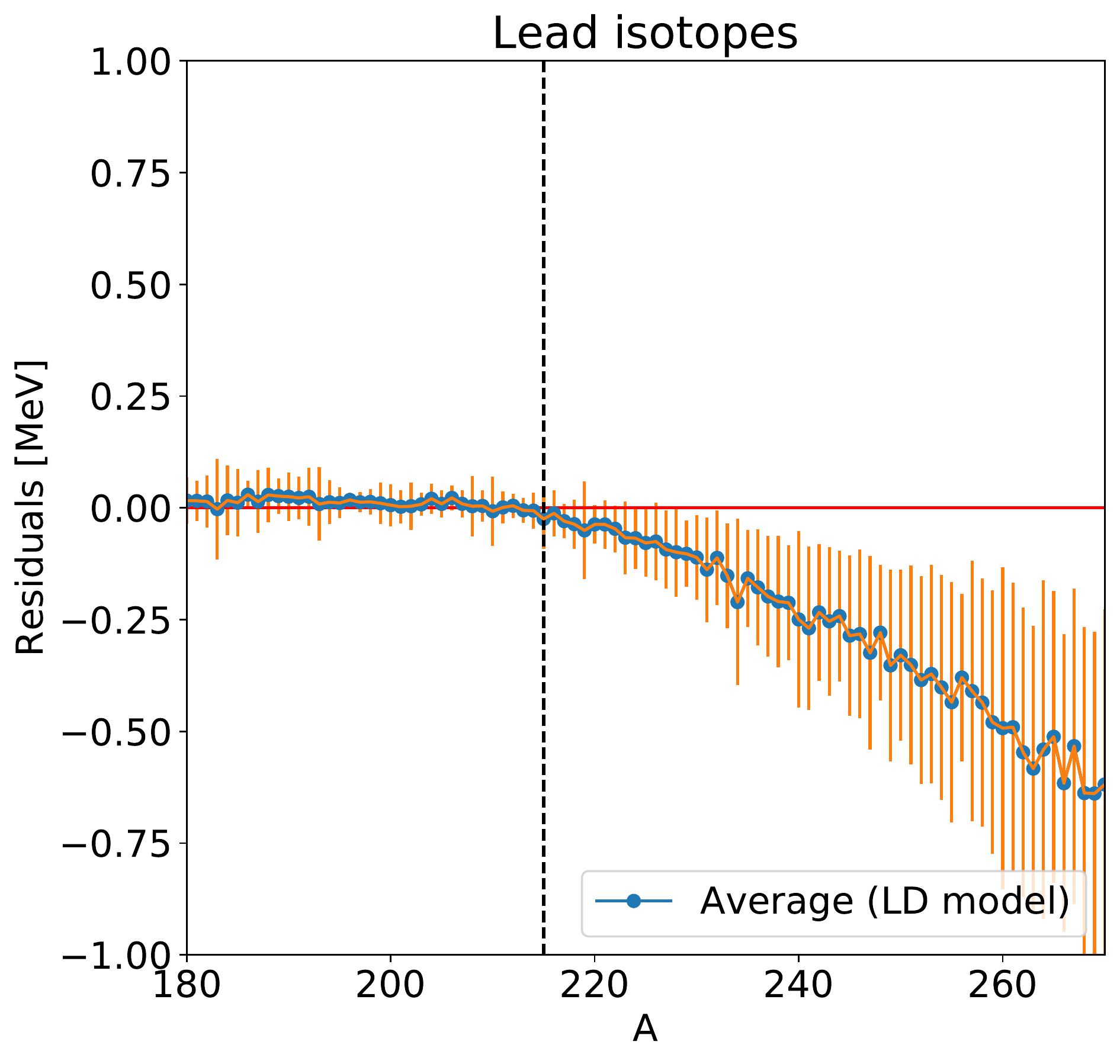}\\
\includegraphics[width=0.43\textwidth,angle=0]{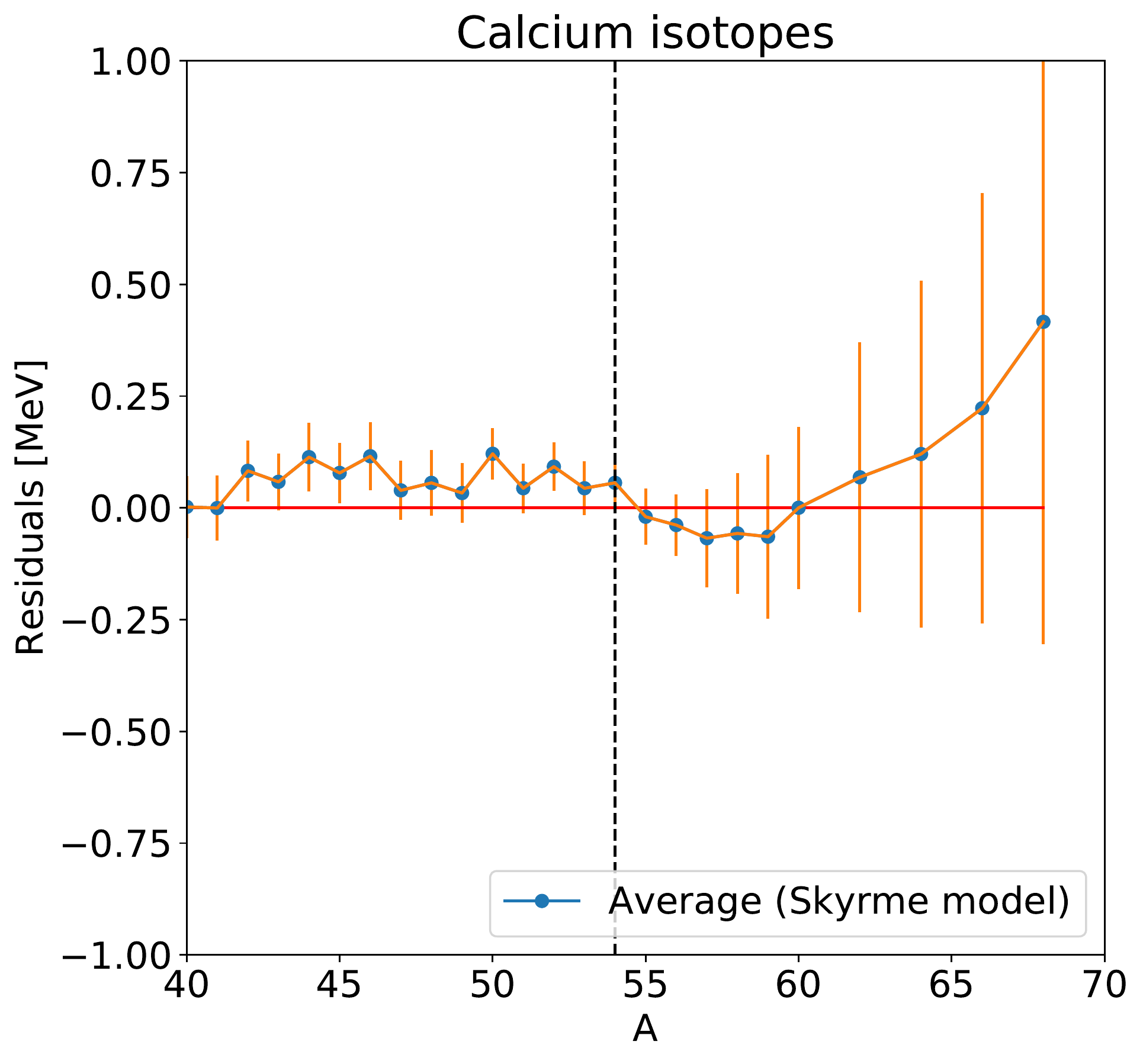}
\includegraphics[width=0.43\textwidth,angle=0]{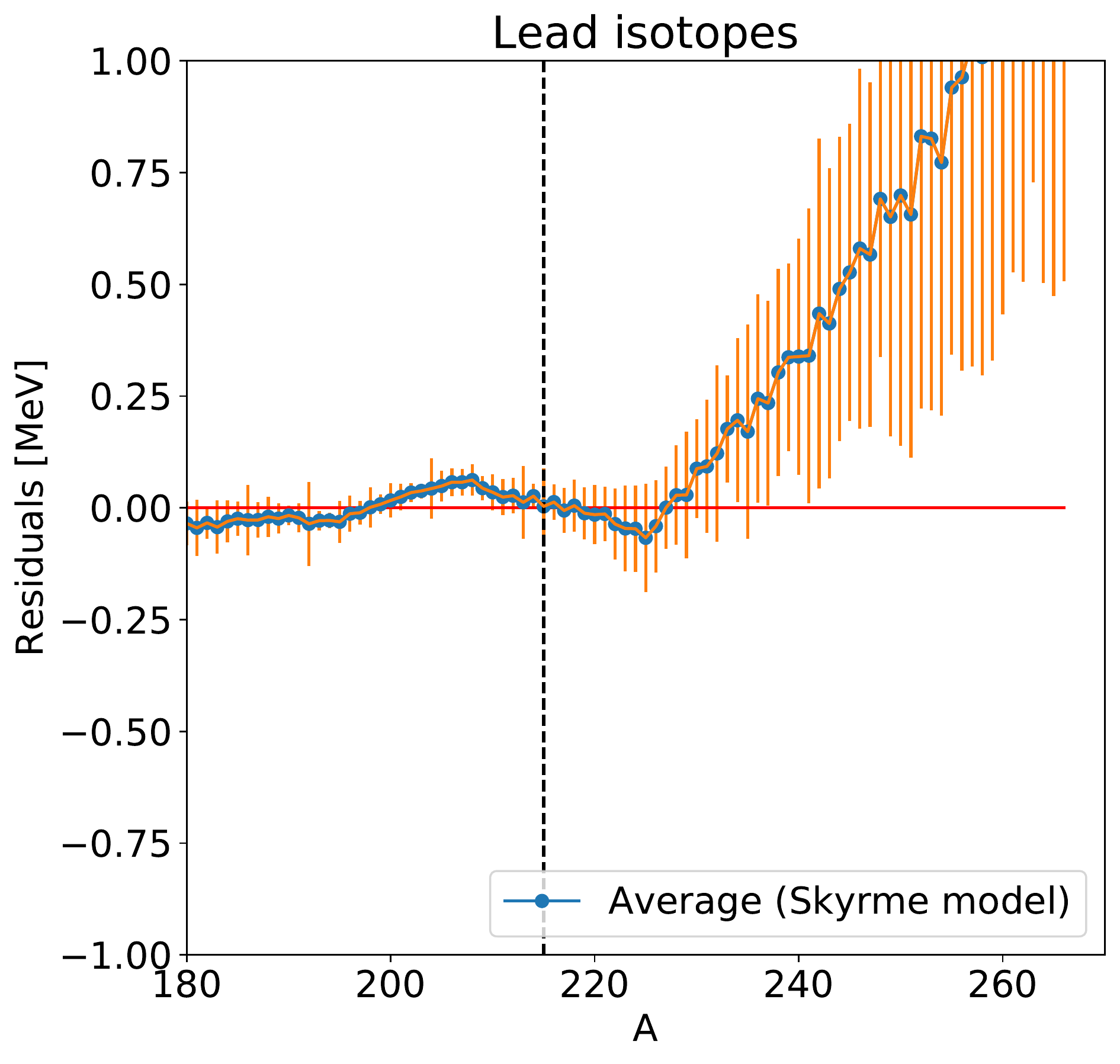}\\
\end{center}
\caption{Same as Fig.\ref{residue}, but using dropout to estimate error bars. See text for details.}
\label{dropmass}
\end{figure}

The result is reported in Fig.\ref{dropmass} for the two data-sets used and for two isotopic chains. The quality of the predictions in the region used for the training is slightly worst, but in the extrapolated region we observe that the behavior of the resulting error bar is somehow intermediate between the epoch-averaging and the bootstrap.

Similarly to the bootstrap case, the use of dropout reduces the dependence of the outcome on the specific choice of training/validation set,  but using much less CPU time. The error bars contain the \emph{true} model only for few isotopes beyond the last known nucleus in the LD model, while for the Skyrme model the error bars do a better job. This may be accidental and not easy to predict without knowing the \emph{true} model.

\section{Conclusions}\label{sec:conc}

We have presented three different methodologies to estimate the error bars of the prediction obtained with a neural network.
Through  a simple toy-example, we have illustrated in detail the calculations of error bars using epoch-averaging, bootstrapping and dropout. We have applied these three methods to the more realistic case of nuclear binding energies. To benchmark the accuracy of the error bars and predictions, we have used synthetic data obtained from two well known models: the liquid drop and the Skyrme SLy4. 
By observing the quality of the predictions and the structure of the error bars on two representative isotopic chains, we conclude that bootstrap and dropout are robust methods since they do not depend too much on the particular choice of the initial training set as in the case of epoch-averaging.
The bootstrap tends to provide the largest error bars that contains the \emph{true} value of a large set of nuclei, unfortunately these error bars are so large that the prediction itself lacks any relevance.
The dropout seems more promising in providing a more reasonable error bar. Moreover, it has an actual regularisation effect during the training phase thus reducing the difference between the performance on the training and the validation set. The effect during the \emph{prediction} phase is to lead to a distribution for the predictions. This allows an \emph{a posteriori} analysis of the results, and thus an error estimation. This mimic the behavior of a more complex Bayesian Neural Network with a reduced computational cost. 
The optimal dropout rate for training and prediction has not been explored in detail and a relevant study, with some additional programming effort, on the dependence of error bars and, more generally, on the extrapolation performances is required.

Other parameters may also increase the variability of our predictions, as for example the initialisation of the weights. Different choices of weights may lead to slightly different results and they should also be taken into account for a more detailed analysis. 

We consider that these results are a first step towards a Machine Learning model for predicting the nuclear masses based on the available, experimental masses. The key ingredients for these models are robust error estimations and validation schemes as discussed in this paper. 

Another ingredient that is required for good performance and to generalise beyond the known masses is a solid Feature Engineering approach. We illustrated with the toy model how the usage of the right feature can dramatically improve the predictions. While the example may seem artificial, it serves the purpose to express the need to carefully designed features. 
This means that neural network can not replace modelling, but only complement it.

\section{Acknowledgement}
The authors thank R. Lasseri for useful discussions in the initial phase of this work. We also thank M. Shelley and D. Regnier for useful comments. 
This work has been supported by STFC Grant No. ST/P003885/1.

\bibliography{biblio}

\end{document}